\documentclass[]{aastex631}

\usepackage{float}

\def\oi{{{\rm O}\,{\sc i~}}}
\def\oii{{{\rm O}\,{\sc ii~}}}

\def\ovi{{{\rm O}\,{\sc vi}~}}

\def\ovii{{{\rm O}\,{\sc vii}~}}
\def\oviii{{{\rm O}\,{\sc viii}~}}

\def\xmm{{\it XMM-Newton}~}
\def\chandra{{\it Chandra}~}

\def\ls{{_<\atop^{\sim}}}
\def\gs{{_>\atop^{\sim}}}

\def\apjl{{ApJL}}
\def\apjs{{ApJS}}

\def\aap{{A\&A}}

\begin{document}

\title{X-Ray Detection of the Galaxy's Missing Baryons in the Circum-Galactic Medium of L$^*$ Galaxies}

\correspondingauthor{Fabrizio Nicastro}
\email{fabrizio.nicastro@inaf.it}

\author[0000-0002-6896-1364]{Fabrizio Nicastro}
\affiliation{Istituto Nazionale di Astrofisica (INAF) - Osservatorio Astronomico di Roma \\
Via Frascati 33 \\
00078 Monte Porzio Catone (RM), Italy}
\affiliation{Department of Astronomy, Xiamen University, Xiamen, Fujian 361005, China}

\author{Y. Krongold}
\affiliation{Instituto de Astronomia - Universidad Nacional Autonoma de Mexico, Mexico City, Mexico}

%\collaboration{20}{(AAS Journals Data Editors)}

\author{T. Fang}
\affiliation{Department of Astronomy, Xiamen University, Xiamen, Fujian 361005, China}

\author{F. Fraternali}
\affiliation{Kapteyn Astronomical Institute, University of Gronigen, Gronigen, The Netherlands}

\author{S. Mathur}
\affiliation{Astronomy Department, The Ohio State University, Columbus, OH, USA}
\affiliation{Center for Cosmology and Astro-Particle Physics, The Ohio State University, Columbus, OH, USA}

\author{S. Bianchi}
\affiliation{Dipartimento di Matematica e Fisica, Universit\'a degli Studi Roma Tre, Roma, Italy}

\author{A. De Rosa}
\affiliation{Istituto Nazionale di Astrofisica (INAF) - Istituto di Astrofisica e Planetologia Spaziali, Roma, Italy}

\author{E. Piconcelli}
\affiliation{Istituto Nazionale di Astrofisica (INAF) - Osservatorio Astronomico di Roma \\
Via Frascati 33 \\
00078 Monte Porzio Catone (RM), Italy}

\author{L. Zappacosta}
\affiliation{Istituto Nazionale di Astrofisica (INAF) - Osservatorio Astronomico di Roma \\
Via Frascati 33 \\
00078 Monte Porzio Catone (RM), Italy}

\author{M. Bischetti}
\affiliation{Istituto Nazionale di Astrofisica (INAF) - Osservatorio Astronomico di Trieste, Trieste, Italy}
\affiliation{Dipartimento di Fisica, Sezione di Astronomia, Universit\'a di Trieste, Trieste, Italy}

\author{C. Feruglio}
\affiliation{Istituto Nazionale di Astrofisica (INAF) - Osservatorio Astronomico di Trieste, Trieste, Italy}

\author{A. Gupta}
\affiliation{Astronomy Department, The Ohio State University, Columbus, OH, USA}
\affiliation{Columbus State Community College, Columbus, OH, USA}

\author{Z. Zhou}
\affiliation{Department of Astronomy, Xiamen University, Xiamen, Fujian 361005, China}

\begin{abstract}
The amount of baryons hosted in the disks of galaxies is lower than expected based on the mass of their dark-matter halos and the fraction of baryon-to-total 
matter in the universe, giving rise to the so called galaxy missing-baryon problem. The presence of cool circum-galactic matter gravitationally bound to its galaxy’s halo up to distances of at least 
ten times the size of the galaxy's disk, mitigates the problem but is far from being sufficient for its solution. It has instead been suggested, that the galaxy missing baryons may hide in a much hotter 
gaseous phase of the circum-galactic medium, possibly near the halo virial temperature and co-existing with the cool phase.
Here we exploit the best available X-ray spectra of known cool circum-galactic absorbers of L$^*$ galaxies to report the first direct high-statistical-significance {\color{black} (best estimates ranging from $4.2-5.6\sigma$,
depending on fitting methodology)} detection of associated \ovii absorption in the stacked \xmm and \chandra spectra of three quasars.
We show that these absorbers trace hot medium in the X-ray halo of these systems, at logT(in k)$\simeq 5.8-6.3$ K (comprising the halo virial temperature T$_{vir} \simeq 10^6$ K). 
We estimate masses of the X-ray halo within 1 virial radius within the interval M$_{hot-CGM}\simeq (1-1.7)\times 10^{11} (Z/0.3 Z_{\odot}) ^{-1}$ M$_{\odot}$.
For these systems, this corresponds to galaxy missing baryon fractions in the range $\xi_b = M_{hot-CGM}/M_{missing}\simeq (0.7-1.2) (Z/0.3 Z_{\odot})^{-1}$, thus potentially closing the galaxy baryon census in typical L$^*$ galaxies. 
Our measurements contribute significantly to the solution of the long-standing galaxy missing baryon problem and to the understanding of the continuous cycle of baryons in-and-out of galaxies
throughout the life of the universe.
\end{abstract}


\section{Introduction} \label{sec:intro}
The galaxy missing baryon problem is present at all halo scales, from dwarves to massive elliptical galaxies and up to groups and clusters of galaxies (e.g. \cite{McGaugh10}) but the baryon deficit is 
larger for smaller halos. 
Galaxy disks in halos of 10$^{12}$ M$_{\odot}$ host only $\sim 20$\% of the expected baryons \citep{McGaugh10}. However, due to their non-baryonic massive halos, the gravitational pull of galaxies 
extends well beyond their stellar disks, up to distances of at least ten times their size. 

\noindent 
Such large volumes of space surrounding the stellar disks are not empty but are known to host clouds of cool (T$\simeq 10^4$ k) gas, gravitationally bound to the galaxy. As suggested by the extensive 
studies carried out, for the local universe, in the Far-ultraviolet band ($\sim 900-2000$ \AA, FUV, hereinafter) with the Hubble Cosmic Origin Spectrograph (COS, \cite{McPhate00}), this cool 
circum-galactic matter (cool-CGM, hereinafter) may be in photoionization equilibrium with the external meta-galactic UV radiation field in which is embedded (e.g. 
\cite{Berg23,Lehner19,Berg19, Wotta19,Lehner18,Keeney17,Werk14,Fox13,Lehner13,Werk13,Stocke13}, but see also \cite{Bregman18} and references therein for alternative possibilities), and often co-exists 
with higher-ionization gas in a different physical state, probed by Li-like ions of oxygen (e.g. \cite{Stocke13,Tchernyshyov22,Prochaska19,Prochaska11}) and/or neon (e.g. \cite{Burchett19}). 
Under the pure-photoionization equilibrium hypothesis, the cool-CGM may contribute importantly to the galaxy baryon budget: for typical L$^*$ galaxies
\footnote{L$^*$ is the characteristic luminosity above which the number of galaxies per unit volume drops exponentially and, in the local Universe, corresponds to the luminosity of a Milky-Way-like galaxy.}
with halo mass of $1.6 \times 10^{12}$ M$_{\odot}$ and a factor of 4 deficit of baryons (e.g. \cite{McGaugh10}), it may account for up to 50\% of the missing baryonic matter (\cite{Werk14}, but see also \cite{Bregman18}). 

At least 50\% of the galaxy missing baryons, however, remains elusive, and is thought to hide in a hotter phase of the CGM \citep{Wijers20}, possibly at the galaxy virial temperature (i.e. T$_{vir}
\simeq 10^{5.7–6.0}$ K, for halo masses of M$_h \simeq 10^{11.8-12.3}$ M$_{\odot}$ at $z=0$, e.g. \cite{Qu18}). 
Observationally, the presence of this hot phase is currently only hinted through Sunyaev-Zeldovich \citep{Bregman22} and low-resolution X-ray \citep{Das20} measurements of the surroundings of local 
L$^*$ galaxy or pioneering X-ray absorption \citep{Mathur21} studies (see below), and more ubiquitously through absorption by moderately-ionized ions of oxygen (e.g. 
\cite{Stocke13,Tchernyshyov22,Prochaska19, Prochaska11}) and neon (e.g. \cite{Burchett19}). 
At such high temperatures, indeed, hydrogen is virtually fully ionized and so difficult to detect. Therefore, the only available tracers (with typical ion fractions of only a few percent) in the FUV portion of 
the electromagnetic spectrum are the Li-like ions of oxygen (at $z\le 0.7$) and neon (at $z \simeq 0.2–1.3$). However, Li-like ions in the CGM may be produced either in tenuous warm clouds purely 
photo-ionized by the external radiation field or in much hotter, mainly collisionally-ionized, gas (see \S \ref{appsec:ion-fractions}).
Distinguishing between these two possibilities is virtually impossible based on the currently available single ion column density measurements (i.e. without estimates of the gas ionization balance) and
thus only loose lower limits on the temperature and mass contribution of this gaseous CGM component have been set so far (e.g. \cite{Tumlinson11,Chen00}). 
 
A better tracer of gas at T$\simeq 10^6$ K,  is the He-like ion of oxygen, which represent about 90--99\% of its element in T$\simeq 10^{5.6–6.2}$ K gas in collisional-ionization equilibrium (CIE; see \S \ref{appsec:ion-fractions}),
and whose main transitions lie in the soft X-ray band.
Pioneering single-target X-ray spectroscopic studies of these transitions from the halos of optimally-selected Lyman-Limit-Systems (low-ionization HI-metal absorbers with moderate 
column density: $16.2 \le$logN$_{HI}$(in cm$^{-2}$)$\le 19$: LLSs, hereinafter) have indeed been recently performed \citep{Mathur21}, but are hampered by the limited resolution and throughput of 
current X-ray spectrometers, and did not produce conclusive results. Detailed surveys of galaxy–-high-ionization X-ray-absorber associations, comparable to low- or moderate-ionization FUV studies 
like the COS-Halos Survey \citep{Werk13} or the CGM2 Survey\citep{Tchernyshyov22}, will have to wait for the next generation of high-throughput X-ray spectrometers (e.g. the Athena-XIFU \cite{Barret18} 
or Arcus \cite{Smith20}; \cite{Wijers20}). In the meantime, exploiting the richness of the Chandra-LETG (Low Energy Transmission Grating, \cite{Brinkman00}) and XMM-Newton RGS (Reflection Grating Spectrometers, 
\cite{denHerder00}) archives and adopting “stacking” techniques (e.g. \cite{Orsolya19}, \cite{Ahoranta20,Ahoranta21}) to perform spectroscopy of optimally selected targets, is a viable alternative. 

Previous studies pursued this strategy by using as signposts for the X-ray transitions the average redshifts of groups of intervening galaxies, and reported non-detection of hot-X-ray intra-group gas 
\citep{Yao10}. Here, instead, we choose to focus on the redshifts of known intervening cool (LLSs) and warm absorbers already extensively studied in the FUV and on their galaxy-associations.

Throughout the paper, we adopt a standard $\Lambda$CDM cosmology, with the latest parameter values from \cite{Planck18} (from the combined analysis of temperature power spectra, high-multipole polarization spectra and 
lensing). In particular, we use universal baryon fraction $f_b= \Omega_b/\Omega_m =0.157$ (Baryon over total-Matter). We also adopt solar metallicities from \cite{Anders89}. In particular we use [O/H]=-3.07. 
Uncertainties are quoted at 68\% significance ({\color{black} for 1-interesting parameter} throughout the paper, unless explicitly stated. {\color{black} Analogously line Equivalent-Width (EW) versus centroid wavelength statistical
significance contours are displayed for $\Delta \chi^2 = n^2$ ($n = 1, 2, 3, 4, 5, ...$), for an easy comparison with the statistical significance of the line evaluated as the ratio between the line EW and its 1-$\sigma$ statistical
uncertainty. 
These $\Delta \chi^2$ correspond to $(1, 2, 3, 4, 5, ...) \sigma$ contours for 1-interesting parameter (the line EW) and  $\simeq (0.5, 1.4, 2.6, 3.7, 4.7, ...) \sigma$ for 2-interesting parameters (line EW and centroid)}. 
{\color{black} All spectral fitting is performed with the fitting package {\em Sherpa} (part of the {\em Ciao} software \cite{Fruscione06}), by exploiting $chi^2$ statistics.
We look for minima, by using two consecutive methods: the {\em levmar} method in {\em Sherpa} \citep{More78} to look for quick solutions, followed by a second fit with the slower {\em moncar}  method in {\em Sherpa}
\citep{Storn-and-Price97} to refine the best-fitting parameters, or look for alternative solutions.}

\section{Sample Selection and the X-ray Halo} \label{sec:xray-halo}
We select as optimal targets the 30 background quasars of \cite{Lehner13} and \cite{Prochaska19} for which LLSs are reported, often associated with moderate-ionization OVI absorbers 
\citep{Fox13,Lehner13}. 

Eleven of these objects have \xmm RGS data available, and two of these have also \chandra-LETG data (see \S \ref{appsec:sample-selection}). {\color{black} Of these eleven targets, we selected only those 
  (a) whose intervening LLSs have been confidently associated to $\sim$L$^*$ galaxies and, among those,  (b) whose total RGS and LETG spectra have signal-to-noise per resolution element SNRE$>4$ in the continuum adjacent the
  LLS-frame \ovii K$\alpha$ transition. The second of these two selection criteria  allows for the search of associated \ovii$_{K\alpha}$ absorption in the individual X-ray spectra of the targets  (see \S \ref{appsec:sample-selection} for
  additional details)}. 

This yielded three quasars, namely PG~1407+265 (observed only with \xmm), PKS~0405-123 and PG~1116+215 (observed with both \xmm and \chandra),
whose lines of sight cross low-ionization LLSs and OVI absorbers at $z_{LLS\#1} =0.6828$ (\cite{Wotta19,Fox13,Lehner13}; LLS\#1, hereinafter),  $z_{LLS\#2} = 0.1671$ (\cite{Wotta19,Fox13,Lehner13,Stocke13}; LLS\#2) and
$z_{LLS\#3} = 0.1385$ (\cite{Wotta19,Fox13,Lehner13,Stocke13}; LLS\#3), respectively (see \S \ref{appsec:sample-selection}). 

The three LLSs that we use to build our X-ray halo, and their galaxy associations, have been reported and discussed in several studies 
\citep{Berg23,Wotta19,Keeney17,Fox13,Lehner13, Werk13,Stocke13,Burchett19} and their properties are reported in Table \ref{apptab:xray-halo} of \S \ref{appsec:xray-halo_properties}.  
They all have HI column densities close to the lowest threshold N$_{HI} = 10^{16.2}$ cm$^{-2}$ of the LLS 
definition in \cite{Lehner13} and are seen at impact parameters (i.e. the line-of-sight to galaxy-center projected distance) $\rho > 90$ kpc (Table \ref{apptab:xray-halo} in \S 
\ref{appsec:xray-halo_properties}). This strongly suggests a cool-CGM (and not extended gaseous disk) origin for the HI-metal absorbers observed in these three systems \citep{Bregman18}.
The three LLSs of our sample also have all co-located \ovi absorption that, however, in the pure-photoionization hypothesys for the cool-CGM traced by the LLSs, cannot be entirely physically associated to the 
cool gas (e.g. \cite{Lehner13}). 

The X-ray halo resulting from the $\sigma^i_{OVII}$-weighted averages (see \S \ref{sec:stacked-spectrum} and \ref{appsec:xray-halo_properties}) of the properties of the LLSs and galaxy-associations 
of our sample, is that of an L* galaxy at $<z_{X-ray-halo}> = 0.276$, with a halo mass of $<M$$_h> \simeq 1.2\times 10^{12}$ M$_{\odot}$ and a virial radius $<R$$_{vir}> = R_{200} \simeq 195$ kpc
(the radius at  which the halo density equals 200$\times$ the universe critical density at the given redshift). 
The constructed X-ray line of sight intercepts the X-ray halo at a projected distance $<\rho> = 115$ kpc from its center ($\sim 0.6$ $<$R$_{vir}>$; last row of Table \ref{apptab:xray-halo} in \S
\ref{appsec:xray-halo_properties}). 
Assuming a spherical halo with virial radius $<R$$_{vir}>$, the line-of-sight pathlength through the X-ray halo is therefore $L = 2 \sqrt{<R_{vir}>^2 -<\rho>^2} \simeq 315$ kpc. 

\section{The X-ray Halo Spectrum} \label{sec:stacked-spectrum}
Table \ref{apptab:xspec} of \S \ref{appsec:sample-selection} lists the 3 targets of our X-ray-halo sample and the properties of their X-ray spectra. The last row of Table \ref{apptab:xspec} contains 
the total available X-ray exposure and SNREs (added in quadrature). 

\subsection{\ovii K$\alpha$ absorption along individual sightlines} \label{sec:line-hints}
We first examined each source X-ray spectrum for a signature of \ovii K$\alpha$ absorption (the strongest transition expected in gas at T$\simeq 10^6$ K) at the redshifts of its LLS.
We did this by {\color{black} first modeling each RGS and LETG spectrum within the fitting-package {\em Sherpa} with the simplest possible astrophysically-motivated continuum model, i.e. a
single power-law plus Galactic absorption (the {\em xstbabs} model in {\em Sherpa}, which includes high resolution edge structures for the K-edges of oxygen and neon and the L-edges of iron).
In all cases, a visual inspection of the residuals showed broad-band systematic wiggles, indicating that the single-power-law model is not an adequate description of the targets' continua.
We then added an $n^{th}$-order polynomial function to the power-law with initially all $n>2$ coefficients frozen to zero, and refitted the data. We iterated the procedure  by gradually freeing $n>2$
polynomial orders until residuals appeared flat over the whole RGS and LETG  5--37 \AA\ (observed) bands. Finally,  to search for \ovii K$\alpha$ absorption at the redshift of the LLS}, 
{\color{black} we added a negative unresolved (Full-Width Half Maximim frozen to FWHM=10 m\AA) Gaussian to our best-fitting continuum models, with position allowed to vary within 1 \AA\ from its
expected position  at the LLS's redshift and free negative-only amplitude. 

Table \ref{tab:single-spectra_linepars} summarizes the best-fitting line parameters, as derived by both fitting the individual spectra (first part of the Table) and by joint-fitting the RGS and LETG spectra of the same targets by linking their  
line EWs to the same value (raws 6 and 7 of the Table, where line-centroids are the $\sigma^i_{OVII}$-weighted averages of the line centroids in the RGS and LETG spectra of the targets).
Errors associated to line centroids are Gaussian-equivalent 1$\sigma$ uncertainties (i.e. FWHM$/(2\sqrt{2ln2}$) derived from the distributions of offsets of known Galactic lines in two samples of RGS and
HRC-LETG spectra (see Fig. \ref{appfig:Galactic_OVII-OI_Centroid-Distribution} in \S \ref{appsec:disp-rels}), while errors on EWs are 1$\sigma$ statistical errors from the data.
The last raw of the Table reports the weighted-average line parameters for the X-ray halos, computed by adopting as weights the statistical significance $\sigma^i_{OVII}$ ($i=1,3$) of the lines, in the RGS spectrum of PG~1407+265 (first raw of
the Table) and the jointly-fitted RGS+LETG spectra of PKS~0405-123 and PG~1116+215 (rows 6 and 7 of the Table). The statistical significance of the X-tay halo \ovii K$\alpha$ line is the sum, in quadrature, of the line significance in those
three spectra, and is used to derive the 1$\sigma$ error on the weighted-average EW
\begin{table}[ht!]
\centering
\caption{ \it Best-fitting parameters of the X-ray-halo lines in the single spectra of the three targets}
\vskip 0.1 in
\begin{tabular}{|l|c|c|c|c|}
\hline
X-Ray Spectrum & $\lambda_{LLS-frame}^{\ovii K\alpha}$ & EW$_{LLS-frame}^{\ovii K\alpha}$ & $\Delta v$ & Significance \\
    & (\AA)  & (m\AA) & (km s$^{-1}$) & \\
  \hline
\multicolumn{5}{|c|}{Fits to Individuals Spectra} \\
\hline 
1: PG~1407+265 RGS & $21.59 \pm 0.03$ & $59 \pm 35$ & $-(140 \pm 420)$ & 1.7$\sigma$\\
2: PKS~0405-123 RGS & $21.54 \pm 0.03$ & $15.7 \pm 7.1$ & $-(830 \pm 420)$ & 2.2$\sigma$\\
3: PKS~0405-123 LETG & $21.57 \pm 0.05$ & $69 \pm 25$ & $-(420 \pm 690)$ & 2.8$\sigma$\\
4: PG~1116+215 RGS & $21.54 \pm 0.03$ & $20.8 \pm 8.0$ & $-(830\pm 420)$ & 2.6$\sigma$\\
5: PG~1116+215 LETG & $21.48 \pm 0.05$ & $29.0 \pm 14.5$ & $-(1670 \pm 690)$ & 2.0$\sigma$\\
  \hline
\multicolumn{5}{|c|}{Joint-fits to RGS+LETG spectra with EWs linked to the same value} \\
\hline 
PKS~0405-123 RGS+LETG & $21.56 \pm 0.04$ & $20.5 \pm 7.3$ & $-(560 \pm 560)$  & 2.8$\sigma$\\
PG~1116+215 RGS+LETG & $21.51 \pm 0.04$ & $18.1 \pm 6.5$ & $-(1250 \pm 560)$ & 2.8$\sigma$\\
  \hline
  \multicolumn{5}{|c|}{Weighted averages and coadded significance} \\
  \hline
X-Ray halo & $21.55 \pm 0.04$ & $28.5 \pm 6.6$ & $-(690 \pm 560)$  & 4.3 \\
  \hline
\end{tabular}
\label{tab:single-spectra_linepars}
\end{table}

\noindent
Finally, the three panels of  figure \ref{fig:individual-contours}, show the confidence-level contours of the LLS-frame \ovii K$\alpha$ line EW and centroid seen in the individual \xmm-RGS and \chandra-LETG
spectra of our three targets (see Figure's caption for details).
\begin{figure}[ht!]
\plotone{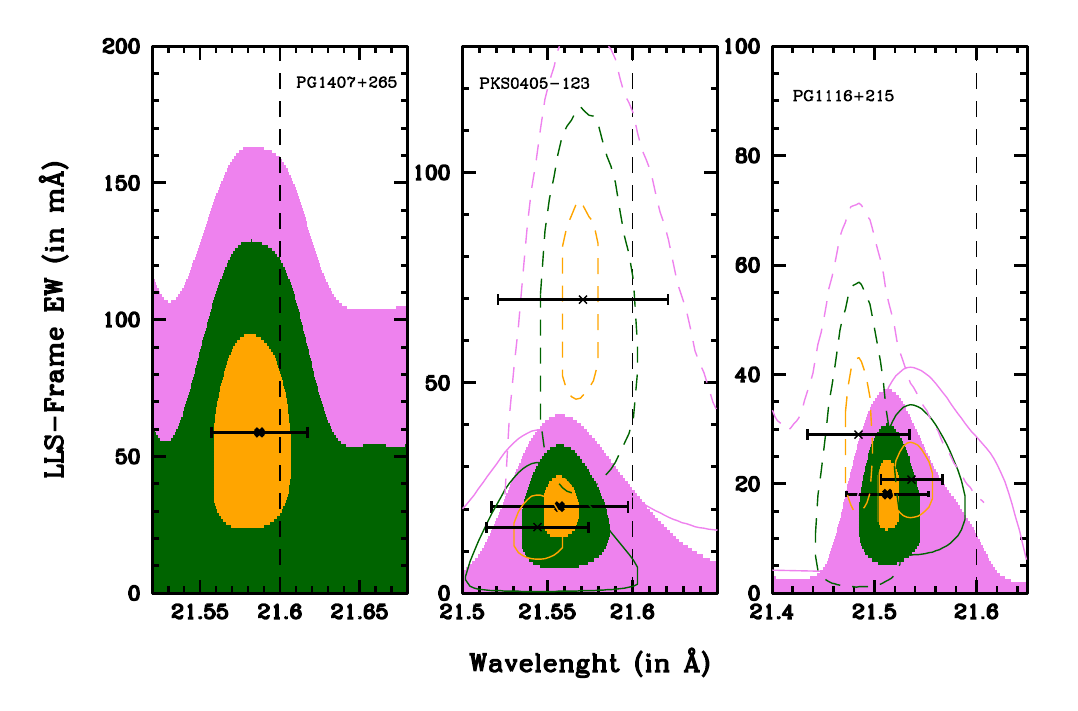}
\caption{1 (orange), 2 (green) and 3$\sigma$ (violet) confidence-level contours of the LLS-frame \ovii K$\alpha$ line EW  and centroid hinted in the individual \xmm-RGS and \chandra-LETG spectra of
our three targets. Colored shaded areas are the EW-$\lambda$ confidence levels for the best-fitting line parameters of the three LLSs in the RGS spectrum of PG~1407+265 (left panel), and the simultaneous RGS+LETG spectra of of 
PKS~0405-123 (middle panel) and  PG~1116+215 (right panel). Solid and dashed curves in the middle and right panels are the confidence level best-fitting line parameters in the individual RGS and LETG spectra of PKS~0405-123 
and PG~1116+215, respectively. {\color{black} Horizontal errorbars are the 1$\sigma$ uncertainties on the X-ray halo line centroids, evaluated as the Gaussian-equivalent standard deviations of the distributions of Fig.
\ref{appfig:Galactic_OVII-OI_Centroid-Distribution} in \S \ref{appsec:disp-rels}}.}
\label{fig:individual-contours}
\end{figure}

Summarizing, none of the single-source spectra shows the clear presence of a possible \ovii K$\alpha$ absorption line imprinted by the halo of the intervening galaxies,
but all of them are consistent with the presence of such a feature at statistical significance levels comprised between $\sigma^i_{OVII} \sim 1.7-2.8$ ($i=1, 5$).
The line hinted in the LETG spectrum of PKS~0405-123 was already reported by our group \cite{Mathur21} at the level of statistical significance shown also here in the middle panel of Fig. \ref{fig:individual-contours} ($\sim 2\sigma$:
dashed contours), and modeled as the hot counterpart of LLS\#2 in association with the OVI absorber reported by \cite{Savage10}. However, the statistical significance of this line alone did not allow us to reach definitive conclusions
on the temperature, column density and mass of this hot-CGM absorber \cite{Mathur21}. 

\subsection{Simultaneous Fit to the \ovii absorbers of the three LLSs} \label{subsec:simfit}
After checking for the presence of X-ray-halo \ovii K$\alpha$ lines in the single \xmm\ and \chandra\ spectra of our targets, we proceded to fit simultaneously the five X-ray spectra of our sample in
the common observed $5-37$ \AA\ band, {\color{black} with the same models used to model the single spectra independently (with continua parameters frozen to their best-fitting values)}, plus the addition of a second negative and
unresolved Gaussian with position linked to that of the first Gaussian through the relative rest-frame positions of the \ovii K$\alpha$ and K$\beta$ transitions, i.e. $\lambda^{OVII K\beta} =  \lambda^{OVII K\alpha} \times (18.63/21.6)$
\footnote{As shown in \S \ref{appsec:disp-rels} (i.e. Fig. \ref{appfig:Galactic_kakb-OI-OII_Delta-Distribution}), linking the positions of the K$\alpha$ and K$\beta$ positions to their expected rest-frame ratio, is extremely conservative
  both for the LETG and the RGS, but allows for a closer comparison of the simultaneous fit to the individual spectra with that performed on the stacked spectrum obtained by rigidly shifting the individual X-ray spectra to either the X-ray- or
  the FUV-LLS redshifts (see \S \ref{subsec:stacked-spectrum}).}
. We did this by exploiting three different methods, namely: (A) leaving all the \ovii K$\alpha$ line positions and EWs and the \ovii K$\beta$ EWs free to vary independently in each spectrum; 
(B) as in A, but linking the K$\alpha$ and K$\beta$ line EWs of the RGS and LETG spectra of the same background targets (PKS~0405-123 and PG~1116+215) to the same values {\color{black} (as expected, if the lines are due to intervening 
hot CGM)}, and (C) as in B, but linking also the line  centroids of the RGS and LETG spectra of the same background targets (PKS~0405-123 and PG~1116+215) to the same values (probably too a strong requirement, given the breadth of the 
distributions of Fig.\ref{appfig:Galactic_OVII-OI_Centroid-Distribution} in \S \ref{appsec:disp-rels}).

\noindent
For each of these methods, A, B and C, the best-fitting line parameters are listed in Table \ref{tab:simfit_linepars}, and the line EW-position confidence-level contours are plotted in the
three panels of Fig. \ref{fig:simultaneous-fit_contours}. Finally, tha raw RGS and LETG data of our targets together with their best-fitting models for the three methods A (oranges olid line), B (green solid line) and C (cyan solid line),
are shown in Fig. \ref{fig:rawdata}. 
\begin{table}[ht!]
\centering
\caption{ \it Best-fitting parameters of the X-ray halo absoprtion lines from the simultaneous fits}
\vskip 0.1 in
\begin{tabular}{|l|c|c|c|c|c|c|}
\hline
Method & $<\lambda_{LLS-frame}^{\ovii K\alpha}>$ & $<\lambda_{LLS-frame}^{\ovii K\beta}>$ & $<$EW$_{LLS-frame}^{\ovii K\alpha}>$ & $<$EW$_{LLS-frame}^{\ovii K\beta}>$ & $\Delta v$  & Significance of \\
    & (\AA)  &(\AA)  &  (m\AA) & (m\AA) & km s$^{-1}$ & the X-ray halo \\
  \hline
  A & $21.54 \pm 0.04$ & $^a$18.58 & $39.1 \pm 7.4$ & $9.4 \pm 5.2$ & $-(830 \pm 560)$ & 5.6$\sigma$ \\
  B &  $21.55 \pm 0.04$ & $^a$18.59 & $28.5 \pm 6.7$ & $16.4 \pm 10.3$ & $-(690 \pm 560)$ & 4.6$\sigma$ \\
  C & $21.55 \pm 0.04$ & $^a$18.59 & $26.4 \pm 7.1$ & $17.3 \pm 8.7$ & $-(690 \pm 560$ & 4.2$\sigma$ \\
  \hline
\end{tabular}

$^a$ Linked to the K$\alpha$ position through the ratio of the rest-frame line positions. 
\label{tab:simfit_linepars}
\end{table}

\begin{figure}[ht!]
\plotone{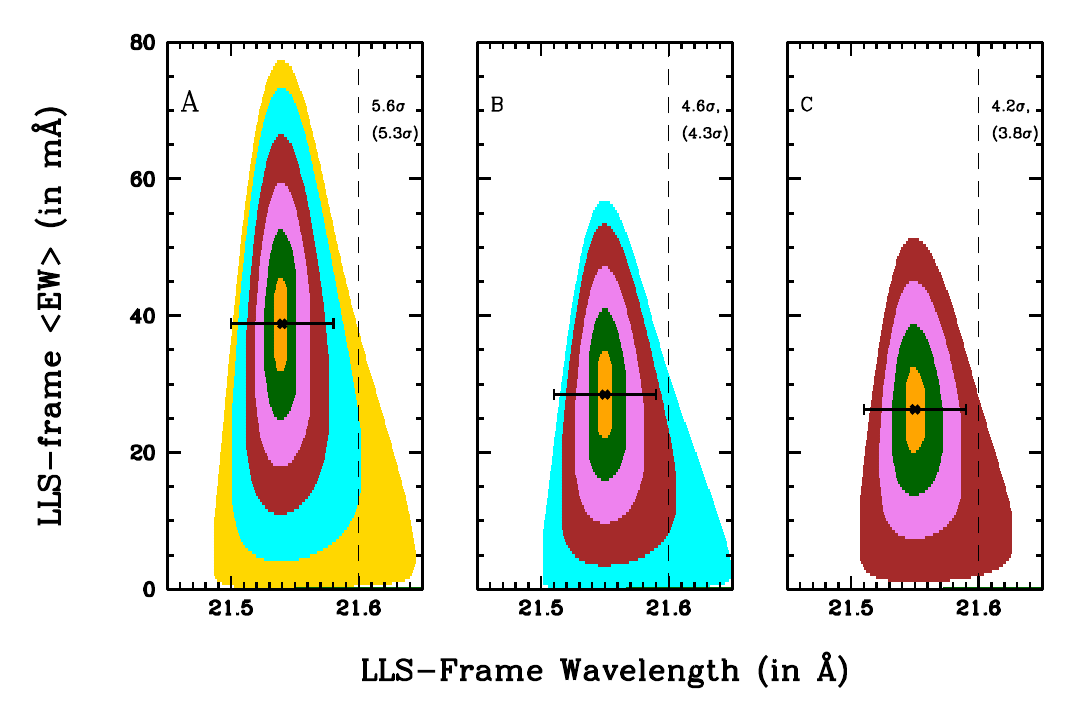}
\caption{\ovii K$\alpha$+K$\beta$ line EW-position confidence-level contours, up to the highest closed contour-level, in the simultaneous fit to the five X-ray spectra of our X-ray halo sample.
  The three panels refer to the three different fitting methods described in the text, and for each method the statistical significance of the highest closed contour is labeled for both one and (two) interesting parameters.}
\label{fig:simultaneous-fit_contours}
\end{figure}

\begin{figure}[ht!]
\plotone{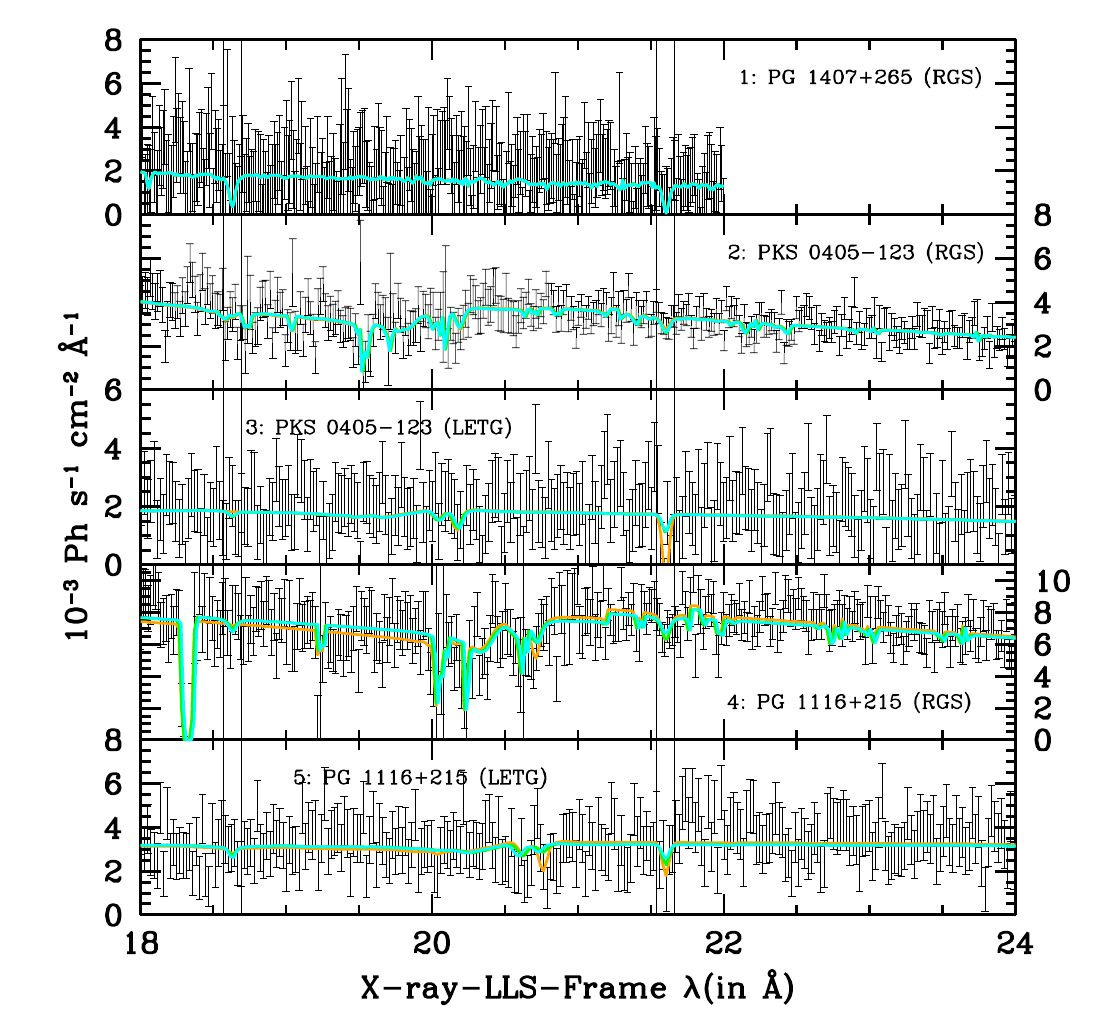}
\caption{Raw \chandra-LETG and \xmm-RGS data of the targets of the X-ray halo, in each X-ray-LLS frame 18-24 \AA\ wavelength range (vertical errobars). In each panel, solid lines are the 
best-fitting continuum plus Gaussian-absorption models for case-A (orange), B (green) and C (cyan), {\color{black} folded through their instrumental responses: the complex and prominent structures visible in the RGS spectra
are instrumental features, due to bad-pixels in the dispersion detectors}. The black rectangles mark the region containing the \ovii K$\alpha$ and K$\beta$ transitions.}
\label{fig:rawdata}
\end{figure}

Uncertainties and confidence-level contours are computed by linking line-centroids and EWs to those of one of the spectra, used as reference.
In particular, we link the \ovii K$\alpha$ line-positions of spectra 2--5 to that of spectrum 1 via the relations $\lambda^{OVII K\alpha}_{i} = \lambda^{OVII K\alpha}_1 \times (\lambda_{bf,i}^{OVII K\alpha}/ \lambda_{bf,1}^{OVII})$ (where the
index $i=2,5$ indicates the $i-th$ spectrum and {\em bf} stays for Best-Fitting).
Similarly, the fluxes $F_i$ of each \ovii K$\alpha$ and K$\beta$ line in spectra 2-5, are are linked to $F_{1}^{K\alpha,K\beta}$ through the relations $F_{i}^{K\alpha,K\beta} = F_{1}^{K\alpha,K\beta} \times (F_{bf,i}^{K\alpha,K\beta}/F_{bf,1}^{K\alpha,K\beta})$. 
This allows us to compute statistical errors on the flux of the \ovii K$\alpha$ and K$\beta$ lines, leaving free to vary independently only three parameters (namely, the \ovii K$\alpha$ line centroid and K$\alpha$ and K$\beta$ 
fluxes of spectrum 1), thus exploiting the combined statistics of all data. Finally, to compute the line position-EW confidence levels of the X-ray halo (i.e. \ovii K$\alpha$+K$\beta$ lines) absorbers plotted in Fig.
\ref{fig:simultaneous-fit_contours}, we also link the flux of  the K$\beta$ line of the reference spectrum 1 to that of its corresponding K$\alpha$ transition, via the relation $F_{1}^{OVII K\beta} = F_{1}^{OVII K\alpha}
\times (F_{bf,1}^{OVII K\beta}/F_{bf,1}^{OVII K\alpha})$, so that the confidence-levels shown in Fig. \ref{fig:simultaneous-fit_contours} are for the combined \ovii K$\alpha$ and K$\beta$ transitions. 
For each of the three cases, the values reported in Tab. \ref{tab:simfit_linepars} for the line centroids and EWs of the two transitions are the $\sigma^i_{OVII}$-weighted averages of their best-fitting values in each spectrum
(Tab. \ref{tab:single-spectra_linepars}), while the statistical significance of the X-ray halo is the coadded (in quadrature) statistical significance of the \ovii K$\alpha$ and K$\beta$ lines (which, by construction, coincides with 
that of the corresponding highest significance closed contour of Fig.  \ref{fig:simultaneous-fit_contours}). 

\subsection{Stacked Spectrum of the X-ray Halo} \label{subsec:stacked-spectrum}
{\color{black} Finally, to fully exploit the whole statistics of the 5 datasets in a single spectrum (combined with the lowest possible number of degrees of freedom)}, we proceeded to blue-shift the five background-subctracted spectra and
their best-fitting continuum models, to both, their own X-ray-LLS redshifts (i.e. the redshift derived from the best-fitting position of the \ovii K$\alpha$ lines in each spectrum: {\em X-ray-LLS spectrum}, hereinafter) and the exact
FUV-LLS redshifts (i.e. the redshifts of the cool-CGM absorbers in the HST-COS spectra of the three targets: {\em FUV-LLS spectrum}, hereinafter), re-grid them over a common $\lambda_{RF} =1-30$ \AA\ (where {\em RF} stays for
Rest-Frame) wavelength grid with bin-size of
30 m\AA\ (about 0.4 and 0.6 times the RGS and LETG LSF-FWHMs, respectively) and stack them together by weighting each spectrum and best-fitting continuum model by its relative signal-to-noise ratio per bin.
{\color{black} Errors on the stacked raw counts per bin were computed in the Poissonian hypothesis ({\color{black} justified by the SNRE$>4$ in the individual X-ray spectra of the taregts of our sample}) as $1 + \sqrt{0.75 + COUNTS}$
  \citep{Gehrels86}. We then ratioed the stacked spectra and their errors with their best-fitting continua to produce the final background-subtracted and continuum-normalized stacked spectra of the X-ray halo.} 

Strong absorption line-like signals are revealed at (left panels of Fig. \ref{fig:stacked-spectrum}) or near (right panels of Fig. \ref{fig:stacked-spectrum}) the rest-frame wavelengths of the strongest K$\alpha$ and K$\beta$ transitions
of the He-like ion of oxygen, both in the X-ray-LLS- (left panels) and FUV-LLS-frame (right panels) continuum-normalized stacked spectra of the X-ray halo. 
To evaluate the centroid positions, equivalent widths (EWs, hereinafter) and statistical significances of these lines, we performed standard spectral fitting of the stacked spectra. We did this by first exploiting the ftools
\citep{Blackburn95} ‘ftflx2xsp’ and ‘ftgenrsp’ to (a) convert the continuum-normalized stacked spectra of the X-ray halo into standard PHA formats and  (b) build an over-semplified  normalized Photon-Redistribution matrix (RSP) with
Gaussians LSF with Full-Width Half Maximum (FWHM) equal to the average RGS and LETG LSF-FWHMs ($\Delta\lambda = 60$ m\AA). The continuum-normalized PHA spectra, folded with their responses, were then fitted in 
{\em Sherpa}, with a model consisting of a constant plus two negative and unresolved (FWHM frozen to 10 m\AA) Gaussians, with all (continuum and lines) parameters free to vary in the fit. 
The fits to both stacked spectra yielded values of the constant fully consistent with unity and visual inspections showed flat residuals over the entire explored band, confirming the accuracy of the continuum modeling of the 5 RGS and LETG
spectra described in \S \ref{sec:line-hints}.
We then froze to constants to 1 and refit the data, obtaining the best-fitting line parameters and statistical significances (i.e. the ratio between the line EW and its
1$\sigma$ statistical error) listed in Table \ref{tab:stacked-spectrum_linepars}, where we also list the 90\% EW upper limit on the H-like oxygen K$\alpha$ transition (forced, in both spectra to have a
frozen line-centroid $\lambda_{RF}^{OVIII K\alpha} = 18.63$ \AA). The top part of the Table is the result of the fit to the X-ray-LLS-frame spectrum, while the bottom part lists the best-fitting parameters obtained on the FUV-LLS-frame spectrum.
\begin{table}[ht!]
\centering
\caption{ \it Best-fitting X-ray halo absorption line parameters}
\vskip 0.1 in
\begin{tabular}{|l|c|c|c|}
\hline
Line Parameter & \ovii K$\alpha$ & \ovii K$\beta$ & \oviii K$\alpha$ \\
  \hline
  \multicolumn{4}{|c|}{X-Ray-LLS Spectrum} \\
\hline
Centroid (in \AA) & $21.604^{+0.007}_{-0.006}$ & $18.64^{+0.08}_{-0.02}$ & $^a$18.97 \\
EW (in m\AA) & $21.6 \pm 3.4$ & $8.6 \pm 4.0$ & $\le 7.9$ \\
  Significance & $6.4\sigma$ & $2.2\sigma$ & 90\%\\
  \multicolumn{2}{|c}{Combined Significance} & \multicolumn{2}{c|}{6.8$\sigma$ }\\
  \hline
  \multicolumn{4}{|c|}{FUV-LLS Spectrum} \\
  \hline
Centroid (in \AA) & $21.54^{+0.02}_{-0.01}$ & $18.647^{+0.009}_{-0.010}$ & $^a$18.97 \\
EW (in m\AA) & $12.9\pm 3.9$ & $12.9\pm 3.8$ & $\le 11.4$ \\
Significance & $3.3\sigma$ & $3.4\sigma$ & 90\%\\
  \multicolumn{2}{|c}{Combined Significance} & \multicolumn{2}{c|}{4.7$\sigma$ }\\
\hline  
\end{tabular}

$^a$ Frozen in the fit. \\
\label{tab:stacked-spectrum_linepars}
\end{table}

Fig. \ref{fig:stacked-spectrum} shows the data and best-fitting model (yellow-curve), of both the X-ray-LLS (left panels) and FUV-LLS (right panels) stacked spectra of the X-ray halo. 
Contour plots of the EW-centroid confidence levels of the two absorption lines, are instead shown in Figure \ref{fig:contours-stacked}. 
Clearly, the \ovii K$\alpha$ and K$\beta$ X-ray halo lines, are present in both the stacked X-ray-LLS and FUV-LLS spectra. The relative position of the best-fitting K$\alpha$ and K$\beta$ line centroids in the FUV-LLS spectrum,
$\Delta\lambda_{K\alpha-Kbeta}^{FUV-LLS}$, is offset from the rest-frame relative position of these transitions by $\sim -1000$ km s$^{-1}$, fully consistent with the observed distributions of \ovii K$\alpha$, K$\beta$ (RGS) and \oi, \oii
K$\alpha$ (HRC-LETG) offsets in the Galactic samples of \cite{Nicastro16a} and \cite{Nicastro16b}, respectively (Fig. \ref{appfig:Galactic_kakb-OI-OII_Delta-Distribution}).
The line EWs measured in the two spectra, are also consistent with each others within their $1-1.5\sigma$ statistical errors, and so are the ratios of the \ovii K$\alpha$/K$\beta$ EWs and therefore the \ovii column densities
(see \S \ref{subsec:ioncolumns}): EW$_{K\alpha}/$EW$_{K\beta} = 2.5 \pm 1.2$ and $1.0 \pm 0.4$, in the stacked X-ray-LLS and FUV-LLS, respectively. 

The \ovii K$\alpha$ line is detected at the highest significance {\color{black} (6.4$\sigma$)}, and exactly at $\lambda_{RF}^{OVII K\alpha} = 21.6$ \AA,  in the stacked X-ray-LLS spectrum.
This is, obviously, by construction, as this spectrum is built by rigidly shiftting, before stacking, each X-ray spectrum to its own X-ray LLS redshift derived from the best-fitting position of the \ovii K$\alpha$ line in each spectrum,
{\color{black} and therefore should not be regarded as an accurate measurement of the actual probability of chance detection of the \ovii K$\alpha$ line in from X-ray-halo (see also \S \ref{subsec:stat-sign} and \ref{subsec:pchance})}.
On the other hand, the statistical significance of the rigidly shifted \ovii K$\beta$ line is higher in the stacked FUV-LLS spectrum (i.e. at the exact FUV LLS redshifts), as it is the the 90\% upper limit on the EW of the \oviii K$\alpha$ transition
at the LLS redshift (Table \ref{tab:stacked-spectrum_linepars}, {\color{black} probably} confirming that RGS and LETG spectra suffer large uncertainties in their dispersion relationship, as demonstrated by the breatdh of the distribution of
Galactic-line relative-centroid offsets in Fig. \ref{appfig:Galactic_kakb-OI-OII_Delta-Distribution}, and suggesting that {\color{black} a more reliable estimate of the \ovii K$\alpha$ and K$\beta$ EWs (and \ovii K$\alpha$ upper limit)
lines from the X-ray halo lies somewhere in between the $4.7\sigma$ and $6.8\sigma$ derived from the stacked X-ray-LLS and FUV-LLS spectra, respectively, and probably closer to the low-boundary of this range}. 

\begin{figure}[ht!]
\plotone{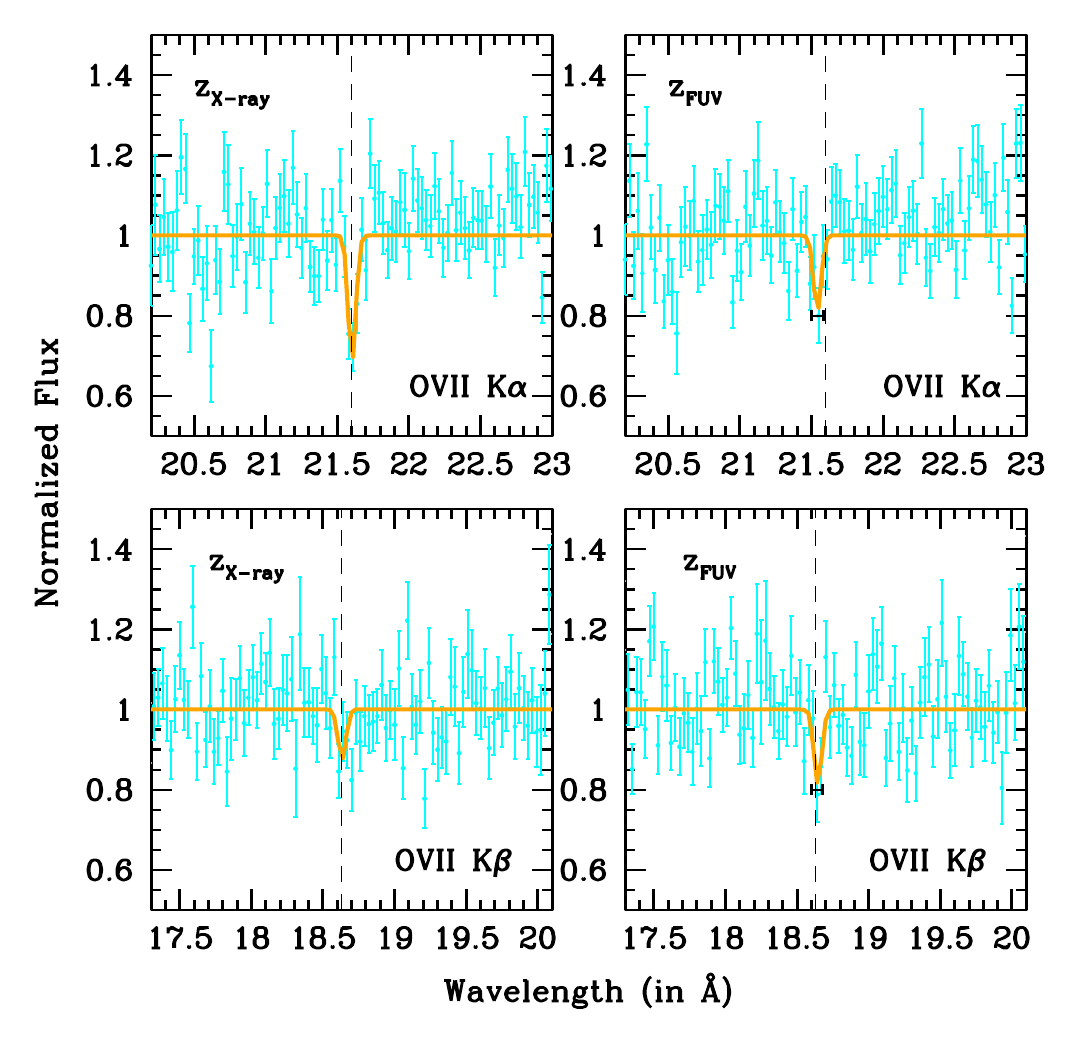}
\caption{Two portions of the stacked X-ray-LLS- (left panels) and FUV-LLS-frame (right panels) spectra of the X-ray halo in continuum-normalized counts and in the wavelength ranges 20.2–23 \AA\ (top panels) and 17.3–20.1 \AA\
  (bottom panels). The spectrum is binned at a resolution of 30 m\AA (about half the LSF-FWHM of both instruments), and has signal-to-noise-per-resolution-element SNRE=29.3 in the continuum at $\lambda = 21.6$.
  The yellow curves are the best-fitting models. The fit yields combined (in quadrature) statistical significances of the \ovii lines of 6.8$\sigma$ (X-ray-LLS spectrum) and 4.7$\sigma$ (FUV-LLS spectrum) (Table
  \ref{tab:stacked-spectrum_linepars})}
\label{fig:stacked-spectrum}
\end{figure}
\begin{figure}[ht!]
\plotone{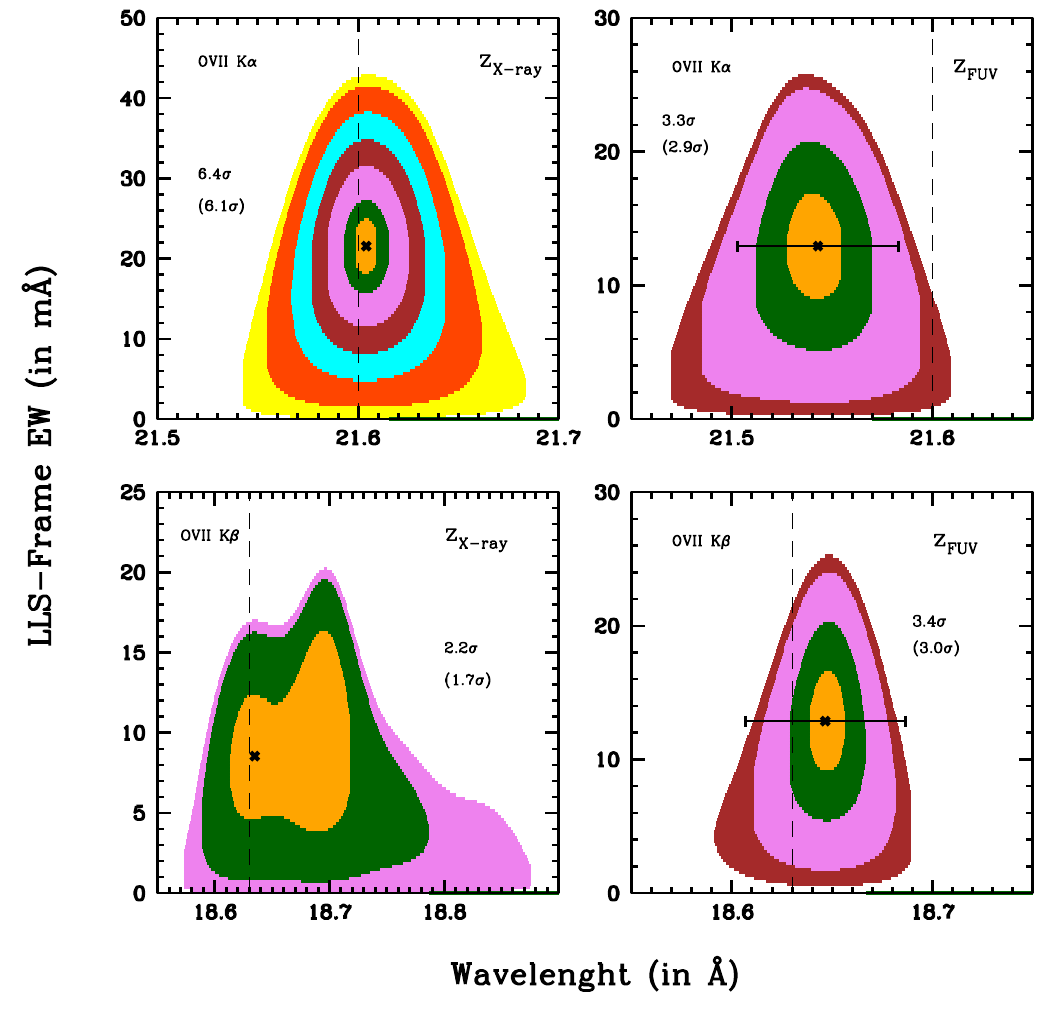}
\caption{\ovii K$\alpha$ (top panels) and K$\beta$ (bottom panels) line EW-position confidence-level contours, up to the highest closed contour-level, in the fits to the continuum-normalized stacked X-ray-LLS (left panels) 
and FUV-LLS (right panels) spectra of the X-ray halo. In each panel, the statistical significance of the highest closed contour is labeled for both one and (two) interesting parameters.}
\label{fig:contours-stacked}
\end{figure}

\subsection{Statistical Significance of the X-ray Halo} \label{subsec:stat-sign}
The total (coadded in quadrature) statistical significance of the \ovii lines of the X-ray halo that we derive from the fitting to the continuum-normalized stacked X-ray-LLS and FUV-LLS spectra are 6.8$\sigma$ and 4.7$\sigma$, respectively, 
{\color{black} The first of these two should, in principle, be compared to the 5.6$\sigma$ statistical significance obtained by fitting simultaneously the 5 X-ray spectra independently with Method A (left panel of Fig.
\ref{fig:simultaneous-fit_contours}), while the second should be compared to either the 4.6 or $4.2\sigma$ obtained through joint-fitting Methods B or C (middle and right panels of Fig. \ref{fig:simultaneous-fit_contours}),
respectively}.

In all cases, difference in statistical significance between join-fitting and stacked-fitting methods, are probably explained by (a) the rigidity of the condition on the relative position of \ovii K$\alpha$ and K$\beta$ lines in each
spectrum (frozen to the rest-frame relative position), which is instead relaxed in the fit to the stacked spectra, where the centroinds of both the K$\alpha$ and K$\beta$ Gaussians are left free to vary in the fit (which yields to a relative
offset of the K$\alpha$--K$\beta$ transitions of about +10 m\AA\ and 60 m\AA\ for the stacked X-ray-LLS and FUV-LLS spectra, respectively, see Table \ref{tab:stacked-spectrum_linepars}), and (b), most importantly, the over-semplified
Gaussian-shaped LSF assumed to build the average RGS+LETG response that we use to fit the stacked spectra (see \S \ref{subsec:stacked-spectrum}): indeed, while a Gaussian-shaped LSF is an excellent approximation of the HRC-LETG LSF,
{\color{black} the RGS LSF is better approximated by a Lorentzian, with broad wings due to electron scattering of photons from the reflection gratings to the dispersive detectors}.
The actual statistical significance of the \ovii signal from the X-ray halo, lies thus somewhere between 5.6--6.8$\sigma$ (joint-fitting case-A vs stacked X-ray-LLS fitting) or 4.2--4.7$\sigma$ (joint-fitting case-C vs stacked
FUV-LLS fitting), {\color{black} and it is probably best estimated by the $4.6-4.7\sigma$ significances derived through the joint-fitting case-B or the fitting to the stacked FUV-LLS spectrum
(1-sided chance detection probabilities of $1.3-2\times 10^{-6}$, increasing to $0.8-1.2 \times 10^{-5}$ after conservatively allowing for redshift trials: see \S \ref{subsec:pchance})}. 

\subsection{Probability of Chance Detection/Identification of the X-ray Halo} \label{subsec:pchance}
Strong intervening \ovii Ka absorbers are rare. Recent hydrodynamical simulations predict that a Universe's random line of sight intercepts 0.17 \ovii Ka absorbers per unit redshift with rest-frame EW$\ge 18$ m\AA\ (250 km s$^{-1}$
at 21.6 \AA; e.g. Figure 6, left panel in \cite{Wijers19}).
Such strong absorbers are practically all in halos (Fig. 6, top-right panel in \cite{Wijers20}), but only about half of these absorbers come from halos with mass between $10^{12}-10^{12.5}$ M$_{\odot}$
(the range of masses of the galaxy’s halos associated to our 3 LLSs with galaxy-association). 
Chance probabilities of expecting 0.17$\times z$ such absorbers and seeing 1, up to the redshifts of our background quasars,
are thus 0.015, 0.043 and 0.043, for PG~1116+215, PG~1407+265 and PKS~0405-123, respectively. Analogously the probability of 
seeing none up to the redshift of PG~1216+069 (for which no galaxy-association is reported), out of any mass halo, is 0.95. 
Finally, then, chances of seeing 1 \ovii K$\alpha$ system with $EW\ge 250$ km/s (rest-frame) along 3 out of the 4 lines of sight whose X-ray spectra are sensitive to such EWs at $>1.5\sigma$, is P$_{Theory}(3|4) =
0.015\times 0.043\times 0.043\times 0.95 = 2.6\times 10^{-5}$, which is to be excluded at $\ge 4.2\sigma$ confidence. 

\noindent
This is the chance-identification probability to see the system at any redshift in the allowed intervals.
Here, instead, we use the FUV-LLS redshifts as priors, so the chance-identification probability should be further weighted by the chance-detection probability of the X-ray lines computed by accounting for a number of redshift trials around
the expected FUV-LLS line position that allows for at least the observed X-ray-LLS and FUV-LLS redshift offsets. 
To be extremely conservative we allow for 3 resolution elements ($>1.25\times$ the maximum observed offset in the HRC-LETG spectrum of PG~1116+215) and an oversampling of each resolution element by a factor of 4, i.e. a total of 12
trials per target-spectrum. Fluctuations, however, could be either positive or negative, in equal number. So, the number of redshift trials should be divided by 2 when assessing the significance of absorption-only lines (or, indifferently, the
chance probabilities computed 1-sided and not 2-sided): that is a total of 6 trials for each of the three targets, given our priors. Our three \ovii K$\alpha$ absorbers are seen at statistcial significances of 1.7, 2.8 and 2.8$\sigma$ in the
X-ray spectra of our targets (\S \ref{sec:line-hints}), implyig associated $z-$trial chance-detection probabilities P$_{z-trial}(1.7\sigma) = [1 - P_{Gauss}(1.7\sigma)]\times (12/2) = 0.53$, P$_{z-trial}(2.8\sigma) = 0.031$ and P$_{z-trial}(2.8\sigma)=0.031$
The lines of sight are all independent, so the chance probability of seeing 3 \ovii K$\alpha$ lines out of 3 (the fourth targets lacks the important prior of galaxy association) within the expected redshifts $\pm$ observed offset,
is the product of the three: P$_{z-trial}(3|3) = 0.53\times 0.031\times 0.031 = 0.00051$, to be exluded at $\ge 3.5\sigma$ (which would raise to $\ge 4.4\sigma$ if also the K$\beta$ lines were considered). 

Finally, then, the chance of detecting the three \ovii K$\alpha$ lines at their statistical-significances and down to the observed EW (or column density), and that these are not associated to hot-gas in the three galaxy-association 
halos, is given by P$_{Theory}(3|4)\times P_{z-trial}(3|3) = 2.6\times 10^{-5} \times 0.00051 = 1.3\times 10^{-8}$, which can be excluded with a Gaussian-equivalent statistical significance $\ge 5.6\sigma$. 
In such an unlikely event, the \ovii K$\alpha$ lines seen in the spectra of our three targets at redshifts consistent (or marginally consistent) with those of the LLSs and their galaxy associations, if real, would have to be imprinted by either
diffuse Warm-Hot Intergalactic Medium (WHIM) gas or hot galaxy halos different from those associated to the LLSs but, in either cases at redshifts very close to those of the three LLSs.

\section{Discussion} \label{sec:discussion}
The estimate of the line EWs from highly-ionized oxygen in the spectrum of the X-ray halo, possibly at least partly associated to the moderately-ionized oxygen seen at the 
LLS redshifts in the FUV spectra of the targets of our sample (Table \ref{apptab:xray-halo}; \cite{Fox13}), allows us to assess the physical state of the hot-CGM in the X-ray halo. 

{\color{black} The \ovii K$\alpha$ and K$\beta$ EW ratios in the stacked X-ray-LLS and FUV-LLS spectra are consistent with each others within their 1$\sigma$ statistical uncertainties, and so are the implied \ovii columns. 
These ratios amount to EW$_{K\alpha}/$EW$_{K\beta} = 2.5 \pm 1.2$ and $1.0 \pm 0.4$, in the stacked X-ray-LLS and FUV-LLS spectra, respectively, and are significantly smaller than the expected optically-thin ratio (i.e.
EW$_{K\alpha}/$EW$_{K\beta} = (f_{K\alpha}/f_{K\beta}) \times (\lambda_{K\alpha}/\lambda_{K\beta})^2 \simeq 6.4$, where $f$ are the oscillator strengths of the transition), suggesting a high degree of saturations of the lines and so a
relatively large column density and/or small Doppler parameter. In the following we derive estimates of the \ovii (and upper limits on the \oviii) column density through the X-ray halo, by also exploiting the constraints on the
weighted-average \ovi Doppler parameter and column density.}

\subsection{Ion Column Densities and Doppler parameter of the Hot-CGM in the X-ray Halo} \label{subsec:ioncolumns}
The resolution of the current X-ray spectrometers is not sufficient to resolve the X-ray-halo lines, thus the estimate of the ion column densities N$_{ion}$ and Doppler parameters $b_{ion} = \sqrt{2kT/m_{ion} + 
\sigma_{turb}^2}$ (where T is the electron temperature of the gas, $m_{ion}$ the ion mass, $k$ the Boltzmann constant and $\sigma_{turb}^2$ the line-of-sight gas turbulence) must rely on the exploitation of 
curve-of-growth (CoG) techniques {\color{black} (e.g. \cite{Nicastro02,Nicastro05,Williams05})}. 
We used our accurate Voigt-profile routines \citep{Nicastro99} to produce a number of CoGs for each of our two \ovii transition and for ranges of values of logN$_{OVII}$(in cm$^{-2}) = 12–19$ and $b_{O}=16-200$ km s$^{-1}$
(the low boundary being set by imposing a minimum gas temperature of T$\ge 2.5\times 10^5$ K, needed to start producing sensible fractions  –- $\ge 0.1$ –- of He-like oxygen in CIE gas (see Fig. \ref{appfig:ion-fractions} in \S
\ref{appsec:ion-fractions}), and unlikely absence of line-of-sight turbulence motion –- $\sigma_{turb}^2=0$), and searched for the N$_{OVII}$--$b_{O}$ solutions that matched our EW measurements, for both
  the X-ray-LLS and FUV-LLS stacked spectra.  
{\color{black} These are shown as orange and green light-shaded regions in Fig. \ref{fig:ioncolumns}, respectively. We find the following two broad (and similar) ion column density intervals 
logN$_{OVII}$(in cm$^{-2}) \simeq 15.9–18.5$, $b_{O} > 16$ km s$^{-1}$ and logN$_{OVII}$(in cm$^{-2}) \simeq 16.4–18$, $b_{O} = 20-88$ km s$^{-1}$, in the X-ray-LLS and FUV-LLS spectra, respectively}. 
\begin{figure}[ht!]
\plotone{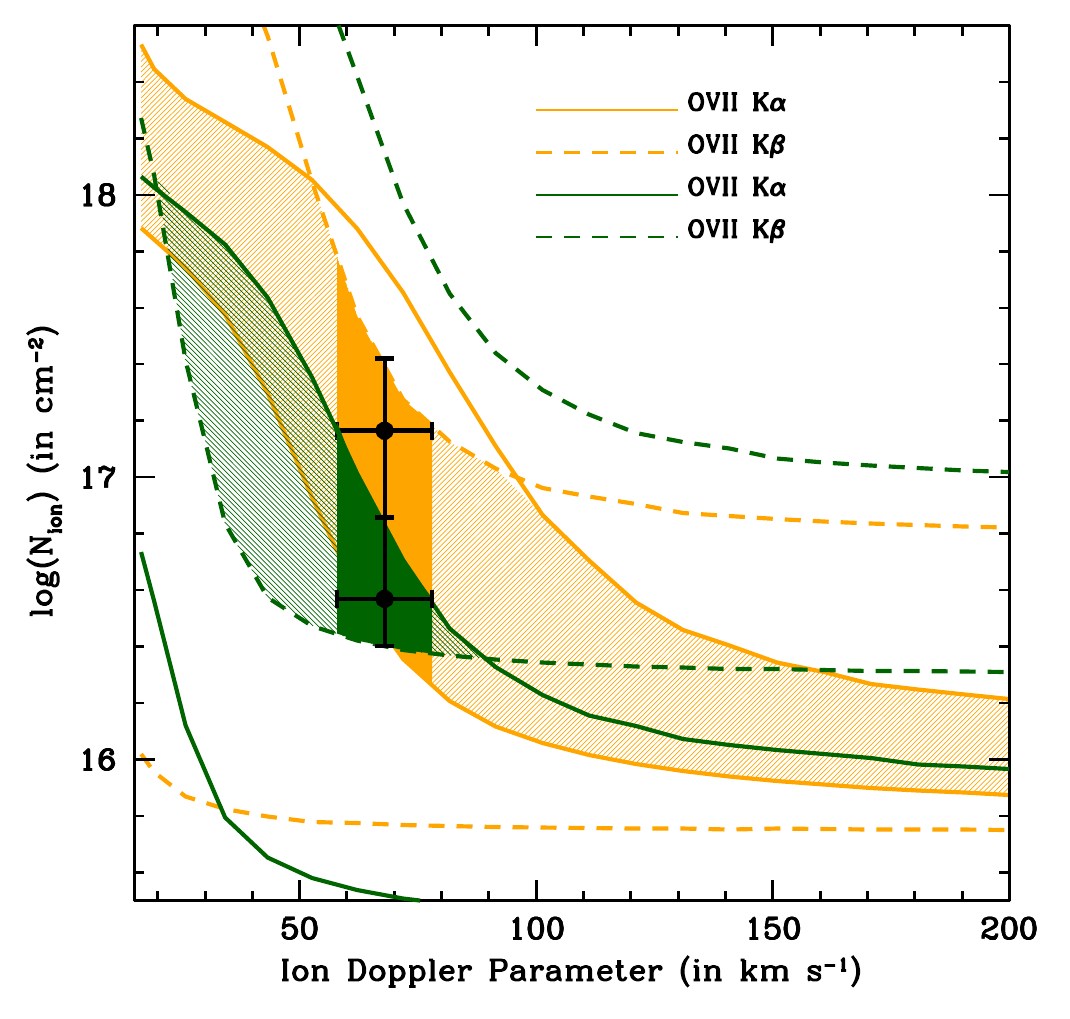}
\caption{Solid and dashed curves show, respectively, the 90\% logN$_{OVII}–b$ contours as derived from the \ovii K$\alpha$ and K$\beta$ transitions measured in the stacked X-ray-LLS (orange) and FUV-LLS (green)
  spectra of the X-ray halo. Light-shaded orange and green regions show the 90\% logN$_{OVII}–b$ solutions for the two spectra, further reduced into the thick-shaded orange and green regions by the fiducial weighted-average value
  $b_{O}=68\pm 10$ km s$^{-1}$ measured for \ovi in the FUV spectra of our sample (black points with error-bars). In this $b_{O}$ interval, and both in the X-ray-LLS and FUV-LLS spectra, the 90\% low-boundary constraint on the K$\alpha$
  (X-ray-LLS spectrum: lower solid orange line) or K$\beta$ (FUV-LLS spectrum: lower dashed green line) transition, sets solid and similar floors to the column density of \ovii through the X-ray halo, at a distance of 115 kpc from the galaxy’s center.}
\label{fig:ioncolumns}
\end{figure}

The range of Doppler parameter values that we measure for oxygen is consistent with many of the LSS \ovi Doppler parameters reported in the literature, and in particular with those of our
3 LLSs ($b_{OVI}^i=28$, 78 and 47 km s$^{-1}$, for LLS\#1, LLS\#2 and LLS\#3, respectively; \cite{Fox13}). If only due to thermal motion (i.e. $\sigma_{turb}^2=0$) $b_{OVII} \sim 20-88$ km s$^{-1}$ (as conservatively measured in
the FUV-LLS spectrum) would correspond to temperatures in the interval T$\simeq (0.25–7.5)\times 10^6$ K.
The exact value of $b$ in this broad interval, however, is not critical with respect to the minimum ion column densities (and so mass of the X-ray halo) allowed by the X-ray data.
Fig. \ref{fig:ioncolumns} shows that in both the X-ray-LLS and FUV-LLS spectra the 90\% low-boundary constraint on the K$\alpha$ (X-ray-LLS spectrum: lower solid orange line) or K$\beta$ (FUV-LLS spectrum: lower dashed green line)
transition, sets stringent lower boundaries to the He-like oxygen column density through the X-ray halo, at a distance of 115 kpc from the galaxy’s center. These are virtually independent on the Doppler parameter in the ranges
$b_{OVII} \gs 50, 100$ km s$^{-1}$ (depending on whether the X-ray-LLS or FUV-LLS solutions are considered; optically-thin limit), whereas $16 \ls b_{OVII} \ls 50, 100$ km s$^{-1}$ (optically-thick regimes) would imply even higher
\ovii column densities. 
{\color{black} In the following we assume that the LLS-associated \ovi absorbers seen in the FUV spectra of our targets are imprinted at least partly by the X-ray halo. Part of the FUV-detected \ovi could belong to a different 
  phase (e.g. \cite{Ahoranta21}), but, given the non-detection of \oviii in the X-ray data (up to the 90\% upper limits listed in Table \ref{tab:stacked-spectrum_linepars}), it is reasonable to assume that at least part of it, is produced by the
  \ovii-bearing phase (see \S \ref{subsec:t-and-nh} and \ref{appsec:ion-fractions}).
  Accordingly, we consider the measured weighted-average \ovi colum (plus its 90\% error) as an upper limit for the \ovi column through the X-ray halo (see \S \ref{subsec:t-and-nh}), and estimate \ovii column densities at
  $<b_{OVI}> = 68 \pm 10$ km s$^{-1}$, the ($\sigma^i_{OVII}\times$N$^i_{OVI}$)-weighted average of the three \ovi absorbers. This corresponds to the X-ray-halo virial temperature logT(in K)$\simeq 6$ (see below) for internal line-of-sight turbulence
  $\sigma_{turb}^2\simeq 48-71$ km s$^{-1}$.

\noindent
In the following we assume the two thick-shaded orange and green regions of Fig. \ref{fig:ioncolumns}  as ranges of \ovii column densities allowed, respectively, by the X-ray-LLS and FUV-LLS spectra: N$_{OVII}^{X-ray-LLS}
\simeq (1.9- 60) \times 10^{16}$
cm$^{-2}$ and N$_{OVII}^{FUV-LLS} \simeq (2.4- 16) \times 10^{16}$ cm$^{-2}$. We also consider the 90\% upper limits on the EW of the \oviii K$\alpha$ transition, inferred by the data at $<b_{OVI}> = 68$ km s$^{-1}$
(Table \ref{tab:stacked-spectrum_linepars}), which yield 90\% \oviii columns N$_{OVIII}^{X-ray-LLS} \le  10^{16}$ cm$^{-2}$ and N$_{OVIII}^{FUV-LLS} \le  2.3\times 10^{16}$ cm$^{-2}$, in the X-ray-LLS and FUV-LLS spectra, respectively.} 

\subsection{Temperature and Equivalent-Hydrogen Column Density} \label{subsec:t-and-nh}
The virial temperature of a $z=0.276$ halo with $<$M$_h> = 10^{12.1}$ M$_{\odot}$ and $<$R$_{vir}> = 195$ kpc (Table \ref{apptab:xray-halo}), is $<$T$_{vir}> = \frac{\mu_p G <M_h> m_p}{2k<R_{vir}>} \simeq 10^{6}$ K \citep{Qu18},
where $\mu_p=0.59$ is the average weight per particle for a fully ionized gas. At this temperature, He-like oxygen largely dominates the ionic abundance distribution of oxygen in CIE gas (left panel of Fig.
\ref{appfig:ion-fractions} in \S \ref{appsec:ion-fractions}), with the H-like and Li-like ions being only $\simeq 2$\% and $<1$\% of the total, respectively.
Li-like oxygen can be more efficiently produced in either CIE gas with T$\ls 4\times 10^5$K (close to the lowest considered value T$=2.5\times 10^5$ K for the \ovii-bearing hot-CGM) or
in low-density (n$_b < 10^{-4}$ cm$^{-3}$) gas photoionized by the external meta-galactic radiation field (left and right panels of Fig.\ref{appfig:ion-fractions} in \S \ref{appsec:ion-fractions}). 
Thus, the observed \ovi could at least partly belong to CGM phases different from the \ovii-bearing hot-CGM phase, including the possibly photoionized cool-CGM.
We therefore conservatively infer temperature and hydrogen-equivalent column density of the hot gas permeating the X-ray halo at an average projected distance of 115 kpc from the galaxy center (about 0.6$\times$ the virial radius 
of our X-ray halo), by combining all the available FUV and X-ray ion column density constraints, but treating the measured average X-ray-halo \ovi column density (plus its 90\% uncertainty) as an upper limit (shaded green
regions of Fig. \ref{fig:hequivalent-column}). 

Practically, we compute ion-by-ion hydrogen-equivalent column densities N$_H$, by dividing the N$_{OVI}$, N$_{OVII}$ and N$_{OVIII}$ ion column densities by the $\sigma^i_{OVII}$-weighted average metallicity
$<Z>\simeq 0.3$ $Z_{\odot}$ reported for the three cool LLS absorbers of our sample (\cite{Wotta19}; Table \ref{apptab:xray-halo}) and that we use in a parameteric form in the folowing to explicitly allow for possible hot- and cool-CGM 
differences, and by the $f_{OVI}$, $f_{OVII}$, $f_{OVIII}$ and $f_{NVI}$ ion fractions in CIE gas (see details and caveats in \S \ref{appsec:ion-fractions}) in the temperature range logT=5.4-6.6, and searching for common logN$_H$-T solutions.
These are shown as orange shaded areas in Fig. \ref{fig:hequivalent-column}, {\color{black} and span similar ranges in the X-ray-LLS and FUV-LLS spectra. In particular, for the X-ray-LLS spectrum (left panel), we find allowed intervals
  logT $^{X-ray-LLS} $(in K)$=5.77–6.27$ and logN$_H ^{X-ray-LLS}$(in cm$^{-2}$)$=(19.86–20.51) -$log$(Z/0.3 Z_{\odot})$, while the FUV-LLS spectrum (right panel), allows for slightly higher temperatures logT$^{FUV-LLS} $(in K)$=5.82–6.35$ and
  H-equivalent column densities logN$_H ^{FUV-LLS}$(in cm$^{-2}$)$=(19.96-20.59) -$log$(Z/0.3 Z_{\odot})$. In both cases, the allowed temperature intervals, encompass the $<$T$_{vir}> \simeq 10^6$ K virial temperature of the X-ray halo,
  and are set by the intersections of the X-ray constraints on the \ovii column density with the FUV N$_{OVI}$ measurements (considered here as upper limits, green solid curve of Fig. \ref{fig:hequivalent-column}) and the X-ray N$_{OVIII}$
  upper limit (brown solid curve of Fig. \ref{fig:hequivalent-column}), on the lower and upper sides, respectively. 
  More importantly, in both the X-ray-LLS and FUV-LLS spectra, the minimum equivalent-hydrogen column density (and so the amount of hot gas in the X-ray halo) is set uniquely by the, very similar, X-ray constraints on the lowest possible column
  density of \ovii, while its upper boundaries are set again by the FUV N$_{OVI}$ and X-ray N$_{OVIII}$ upper limits.}
\begin{figure}[ht!]
\plotone{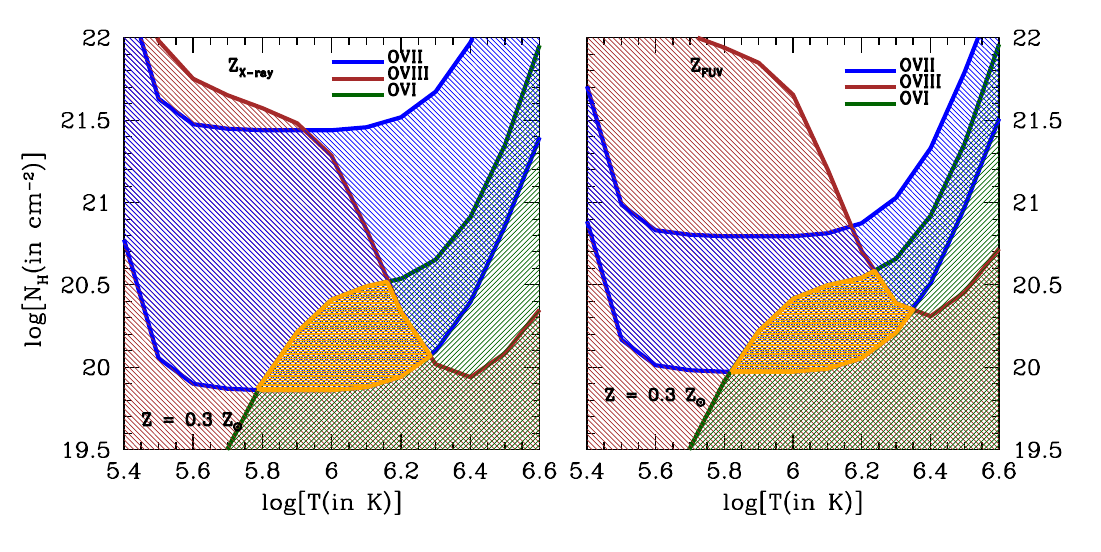}
\caption{Constraints on the H-equivalent column density of the X-ray halo at a projected distance $<\rho>=115$ kpc, obtained by dividing the ion column densities of \ovii (blue curves and shaded area), \oviii (brown curve and shaded area) and
  average \ovi (green curve and shaded area; \cite{Fox13}), at the weighted-average $<b_{OVI}>=68\pm 10$ km s$^{-1}$, by their ion fractions in CIE gas (see \S \ref{appsec:ion-fractions}) and by the average $Z=0.3 Z_{\odot}$ metallicity observed
  for our three cool-CGM systems, as a function of the temperature of the hot-CGM gas.
  The H-like ion of oxygen is not detected in the X-ray-halo spectrum and its column density curve in the figure is a 90\%-confidence upper limit (brown curve and shaded region). 
  \ovi, instead, are detected in the FUV spectra but are considered here as upper limits to allow for at least a portion of this ion to be produced in physical phases different from the \ovii-bearing phase.
  {\color{black} The left panel is for the stacked X-ray-LLS spectrum, while the right panel shows solutions for the FUV-LLS spectrum.}}
\label{fig:hequivalent-column}\end{figure}

Indeed, \ovi alone would favor solutions at temperatures logT(in K)$\simeq 5.4-5.6$ (where the \ovi ion fraction is $\gs 0.1$, in CIE gas: see Fig. \ref{appfig:ion-fractions} in \S \ref{appsec:ion-fractions}),
which would yield equivalent-hydrogen column densities (and so baryonic mass) at least an order of magnitude lower than the boundaries set by the \ovii X-ray measurements in the two stacked spectra of the X-ray halo. 
If only X-ray oxygen data are considered, instead, the low-boundary of the temperature interval is set uniquely by the need of producing detectable fractions of \ovii (i.e. logT(in K)$\gs 5.4$), but the
equivalent-hydrogen column density (and so the baryonic mass: see below) is still lower-bounded at {\color{black} logN$_H ^{X-ray-LLS}$(in cm$^{-2}) \simeq 19.86 -$log$(Z/0.3 Z_{\odot})$ or
  logN$_H ^{FUV-LLS}$(in cm$^{-2}$)$\simeq 19.96 -$log$(Z/0.3 Z_{\odot})$, and allowed to be as large as logN$_H^{FUV-LLS}$(in cm$^{-2})\simeq 20.85 -$log$(Z/0.3 Z_{\odot})$, or implausibly large as 
logN$_H^{X-ray-LLS}$(in cm$^{-2})\simeq 21.45 -$log$(Z/0.3 Z_{\odot})$}.
It is only by combining the FUV and X-ray column density constraints that we can set stringent lower and upper limits to both the temperature and the equivalent-hydrogen column density (and, in turn, mass:
see below) of the X-ray halo.

\subsection{Mass of the Hot-CGM in the X-ray Halo} \label{subsec:mass}
{\color{black} Despite the very similar ranges of temperatures and equivalent-H column densities allowed by either the stacked X-ray-LLS or FUV-LLS spectra, in the following we continue providing separate estimates for the two cases, both for
  the line-of-sight volume density of the hot-CGM and its mass}. 

Our idealized model for the X-ray halo is that of a sphere centered in the galaxy center and filled with 2 gaseous phases (cool and hot, each isothermal) mutually complementing each other spatially.
The hot phase is diffuse and extends from the galaxy center out to the virial radius $R_{vir}$, with density decreasing radially. 
The cool and condensed phase is that of our three LLSs, observed in the FUV through low-ionization metals and H (e.g. \cite{Lehner13}). 

To estimate the average density of the hot-CGM phase of the X-ray halo, we need to estimate the maximum line-of-sight pathelength available for the hot gas, which, under our assumptions, is simply given 
by the total available pathlength covered by the X-ray halo at the projected distance $<\rho>$ minus the thickness of the cool-CGM clouds along the line of sight.  
With the assumed geometry, the pathlength crossed by our lines of sight at a projected distance $<\rho>$ and through the X-ray halo, is simply given by 
$L = 2\sqrt{<R_{vir}>^2 - <\rho>^2}\simeq 315$ kpc. The ratio between the total thickness of the cool-CGM clouds along the line of sight (i.e. the diameter of a single spherical cloud times the number of 
clouds shadowing each other along the line of sight, i.e. \cite{Stocke13}) and $L$, defines the line-of-sight covering factor of the cool phase $f_l^{cool}$.
The total thikness of the cool-CGM clouds has been estimated  for several LLSs by matching the measured ion column densities with predictions by photoionization-equilibrium models in which a halo cloud of gas 
with constant density $n_H^{cool}$ is illuminated by the metagalactic radiation field at the redshift of the LLS (e.g. \cite{Lehner13,Stocke13}). 
For our 3 LLSs \cite{Lehner13} and \cite{Stocke13} derive: N$_H = (10^{20},10^{18.55},10^{19.1})$ cm$^{-2}$ and $n_H = (10^{-3.1},10^{-3.1},10^{-4})$ cm$^{-3}$ for (LLS\#1,LLS\#2,LLS\#3), respectively. 
This gives $\sigma^i_{OVII}$-weighted averages of $<$N$_H>^{cool} = 10^{19.469}$ cm$^{-2}$ and $<n_H^{cool}> = 5.2\times 10^{-4}$ cm$^{-3}$ for the cool-CGM of the X-ray halo, and thus a 
line-of-sight thickness of the clouds of $l(X-ray-halo) \simeq 18$ kpc. This yields a line-of-sight covering factor of the cool-CGM  through the X-ray halo $f_l^{cool} = 0.057$. 

\noindent
Finally, the average density of the hot-CGM phase at a projected distance $<\rho>=115$ kpc through the X-ray halo, is thus given by $<n_H^{hot}> =$N$_H^{hot} / [(1 - f_l^{cool}) L] 
\simeq (8-35)\times 10^{-5} (Z/0.3Z_{\odot})^{-1}$ cm$^{-3}$ (X-ray-LLS spectrum) or $<n_H^{hot}> = \simeq (10-42)\times 10^{-5} (Z/0.3Z_{\odot})^{-1}$ cm$^{-3}$ (FUV=LLS spectrum), and it modulates by a factor of
$n_H^{hot}(@\rho;l=0)/n_H^{hot}(@\rho;l=\pm L/2) \simeq [1 + (L/2\rho)^2]^{3\beta/2} \simeq 3^{3\beta/2}$ from the near-side through the far-side of the halo (here $@\rho;l=$ means equivalent H density at the impact parameter distance
$\rho$ from the center of the galaxy and at line-of-sight length $l$, and $\beta$ is the spectral index of the density profile we adopt to estimate the mass of the hot-CGM: see below). 
The average density is thus only $\ls 7 (Z/0.3Z_{\odot})^{-1}\times$ lower than that estimated for the cool-CGM phase under the pure-photoionization equilibrium and constant gas-density hypothesis. 
This, combined with temperatures of the two phases that differ by factors $\sim 100$, gives pressures that differ by factors $\gs 14 (Z/0.3Z_{\odot})$.  Pressure equilibrium between the two phases would then 
require either $\gs 14 (Z/0.3Z_{\odot})\times$ lower temperatures of the hot phase, inconsistent with the reported detection of \ovii, or $\gs 14 (Z/0.3Z_{\odot})\times$ lower average densities of the hot-phase along 
the line of sight, which, in turn, would require unphysically long line-of-sight pathlentghs of $\gs 4 (Z/0.3Z_{\odot})$ Mpc. 
This suggests that either pressure equilibrium between the two co-existing phases is not at work, or that the cool-CGM clouds are actually $\gs 14 (Z/0.3Z_{\odot})\times$ denser (and so smaller) than inferred under the pure 
photo-ionization hypothesis. {\color{black} In the latter case, the average linear size of the cool-CGM clouds, would be of the order of $1-2$ kpc, i.e. similar to that inferred by the angular-size of typical Galactic HI Compact High-Velocity Clouds (CHVCs)
if at a distance of $\sim 100$ kpc from the Galaxy's center (e.g. \cite{Putman12})}. At such cool-CGM densities, photo-ionization by the external radiation field would be less effective and alternative (or concurring) ionization mechanisms should be
at work (see, e.g. \cite{Bregman18} and references therein), but the cool-CGM clouds could then be pressure-confined by the hot gas (e.g. \cite{Armillotta17,Afruni21} and references therein).
\begin{figure}[ht!]
\plotone{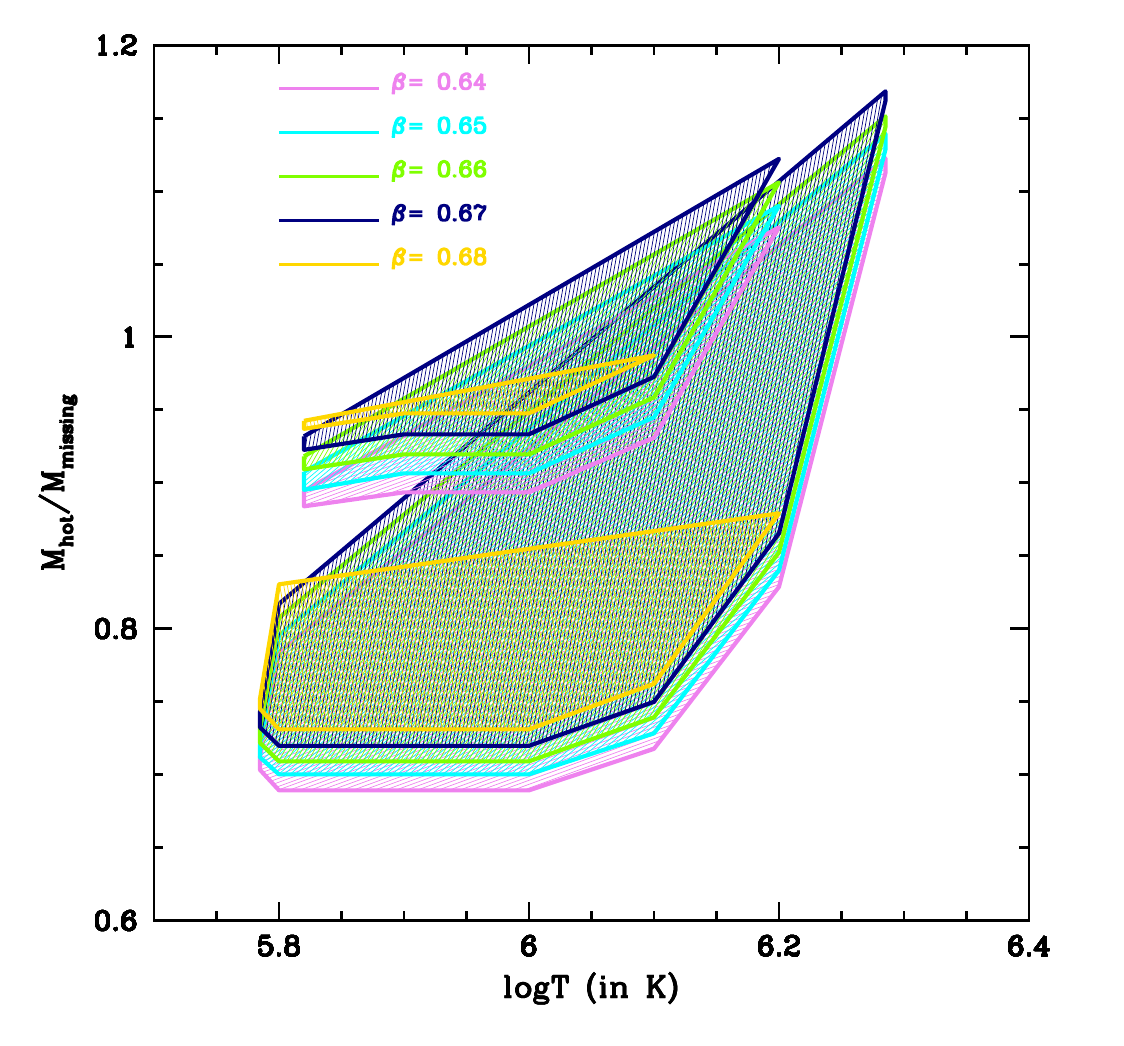}
\caption{Constraints on mass and temperature of the hot-CGM in the X-ray halo, as estimated from the X-ray-LLS (larger polygonal regions) and FUV-LLS (smaller polygonal regions) spectra, for density spectral indeces in the range $\beta = 0.64-0.68$
  (isothermal halo) and hot-CGM metallicity $Z = 0.3 Z_{\odot}$.}
\label{fig:mass}
\end{figure}

To estimate the mass of the hot phase, we assume a volume filling factor $(1 - f_V^{cool}) = (1 - 0.75\times f_l^{cool}) \simeq 0.957$, again complementary to that estimated for the cool-CGM (the factor
0.75 accounts for the occurence of LLS detections in the samples observed with the HST-COS, e.g. \cite{Werk14} and references therein).
For the radial baryon density law of the X-ray halo, we assume a $\beta$-profile \citep{Cavaliere76}: $n_H^{hot}(r) = n_0^{hot} [1 + (r/R_c)^2]^{-3\beta/2}$. 
We integrate the density profile from $r=0$ up to $r=R_{vir}$ for a number of values of the model parameters, searching for those solutions that match the entire range of allowed hot-CGM N$_H$
observed at the average projected distance $<\rho> = 115$ kpc, i.e.: 
\begin{equation} \label{eq:solutions}
N_H = 2 \int_{0}^{L/2} n_H^{hot}(r) dl = 2 \int_{0}^{\alpha_{max}} n_H^{hot}(\alpha;\rho) (\rho/cos^2\alpha) d\alpha, 
\end{equation}
where $dl$ is the increment along the line of sight, $\alpha$ is the angle between the projected distance $\rho$ (i.e. the plane of the sky) and the radius of the halo at a line-of-sight depth $l = 
\sqrt{r^2 - \rho^2} = \rho\ tg(\alpha)$, and $tg(\alpha_{max}) = 2\sqrt{(R_{vir}/\rho)^2 -1}$.

Our X-ray halo has dynamical mass $<$M$_h> \simeq 10^{12.1}$ M$_{\odot}$, stellar mass $<$M$^*>\simeq 10^{10.53}$ M$_{\odot}$ and, under the pure photoionization equilibrium hypothesis (i.a. assuming
an average cloud gas density $<n_H>^{cool} = 5.2\times 10^{-4}$ cm$^{-3}$), cool-CGM gas mass (whithin 1 virial radius) $<$M$_{cool-CGM}>\simeq 10^{10.27}$ M$_{\odot}$ (Table \ref{apptab:xray-halo}). 
This yields a missing baryon mass of M$_{missing} = f_b<$M$_h> - <$M$^*> - <$M$_{cool-CGM}> \simeq 1.45 \times 10^{11}$ M$_{\odot}$ (where $f_b =0.157$ is the universal baryon fraction, \cite{Planck18}). 
By exploring reasonable values of the $\beta$-profile parameters, i.e. $n_0^{hot} = 4\times 10^{-4} - 0.1$ cm$^{-3}$, $R_c = 1-5$ kpc (for reference, central volume density and core radius of the hot halo 
of the Milky-Way have been estimated in the ranges $n_0^{MW} \simeq (0.12-1.2) \times 10^{-2}$ cm$^{-3}$, $R_c=1-3$, \cite{Bregman18}, or $n_0^{MW} \simeq (0.7-6.6) \times 10^{-2}$ cm$^{-3}$, 
$R_c=0.6-2.7$, \cite{Nicastro16a}) and $\beta = 0.64-0.68$ (i.e. centered on the 2/3 value corresponding to an isothermal halo), and accepting only solutions at radii $r/$R$_{vir} = 1.000 \pm 0.005$, we find ranges of allowed
masses M$_{hot-CGM} \simeq (1 - 1.7) \times 10^{11} (Z/0.3Z_{\odot})^{-1}$
corresponding to temperatures in the interval logT(in K)$ \simeq 5.78-6.28$, 
for core radii and volume densities in the ranges R$_c = 3.5-5$ kpc, $n_c^{hot} = 0.02-0.05$ cm$^{-3}$.
  
This mass is at least twice the combined mass of the stellar disk and cool-CGM of the X-ray halo and, more importantly, at least 70\% of the galaxy's missing mass M$_{missing}$. 
The fraction of hot over missing baryons at the halo virial radius lies in the ranges $\xi_b = \frac{M_{hot-CGM}}{M_{missing}} \simeq (0.7-1.2) (Z/0.3Z_{\odot})^{-1}$, 
depending on the exact value of the density-profile spectral index $\beta$ (Fig. \ref{fig:mass}, where smaller regions are estimates from the FUV-LLS spectrum and larger regions from the X-ray-LLS spectrum),
with steeper profiles associated to higher central density, and so mass, solutions.
At the halo virial temperature the fraction of hot over  missing baryons of an isothermal halo ($\beta = 2/3$) lie in the ranges $\xi_b \simeq (0.72 - 1.02) (Z/0.3Z_{\odot})^{-1}$. 

This not only implies that virtually all the baryons that were still missing can now be accounted for by the hot-CGM gas, but has also important consequences for our understanding of the galaxy-CGM
and galaxy-IGM feedback processes throughout the universe lifetime and can help refining feedback prescriptions in hydro-dynamical simulations.
A dense hot virialized CGM containing the vast majority of the expected baryons within its virial radius, suggests that accretion mostly occurred in hot-mode and at the rate given by the cosmological
baryon fraction (e.g. \cite{vandeVoort12}), while the feedback from supernovae and/or past nuclear activity was not sufficiently efficient to expel a significant fraction of the baryons beyond R$_{vir}$.
However, the relative metal richness of the cool- and so probably the hot-CGM (Table \ref{apptab:xray-halo}), may point at an important contribution of feedback (e.g. supernova winds and/or past nuclear activity) 
for its metal pollution. In this scenario, we can speculate that the accretion of fresh gas to feed the star formation in the disc likely takes place via a slow cooling of the hot CGM (e.g. \cite{Fraternali17,Hafen22}).

Our simple spherical, isothermal halo filled with a 2-phase (cool and hot) gas, is clearly an idealization. However, the low-boundary mass of the hot component (M$_{hot-CGM} \sim (1-1.3)\times 10^{11}$ M$_{\odot}$,
depending on whether the stacked X-ray-LLS or FUV-LLS spectrum is considered: Fig. \ref{fig:mass}) is a rather strict and conservative limit, imposed solely by the large amount of \ovii seen in the stacked X-ray spectrum at a projected distance of
$\sim 0.6$ R$_{vir}$. 
Flattening the density profile to $\beta=0.4$ lowers the minimum fraction of missing mass allowed by the solutions by about 15\%, but increases the maximum allowed fraction by a factor of $\sim 3$.
and can easily accommodate missing-baryon masses well within $R_{vir}$. 
Reducing the volume covering factor of the hot component, increasing the hot-CGM metallicity to $>0.3 Z_{\odot}$, modifying the geometry of the halo and/or considering non-equilibrium, multi-temperature,
collisional-ionization models (e.g. \cite{Gnat07}), can help reducing the mass of the \ovii-bearing gas but would still leave large portions of the density-profile parameter space for solutions that close the galaxy
baryon census. 

\section{Conclusions} \label{sec:conclusions}
We reported the first direct detection of \ovii (K$\alpha$ and K$\beta$) absorption in the stacked (or jointly-fitted) X-ray spectra of three LLSs+\ovi absorbers seen in the FUV and associated to the halos of three $\sim$L$^*$ galaxies.
We identify the X-ray absorbers with large amounts of hot gas co-existing with the cool-CGM of these systems and filling their halos. 
In summary, we found that:

\begin{itemize}
\item{} the X-ray halo is detected in the X-ray spectra of the three quasars PG~1417+265, PKS~0405-123 and PG~1116+215 via \ovii K$\alpha$ and K$\beta$ lines, at positions offset by the expected FUV-LLS-frame positions
  by $\sim 300-1600$ km s$^{-1}$, an offset velocity interval consistent with the centroid-offset distributions observed in two samples of Galactic \ovii K$\alpha$, K$\beta$ and \oi, \oii K$\alpha$ absorption lines; 
\item{} the combined (K$\alpha$+K$\beta$) statistical significance of the X-ray halo is comprised between 4.2--5.6$\sigma$, in the jointly-fitted spectra of our five targets (and 4.7--6.8$\sigma$, in the stacked spectra of the
X-ray halo); 
\item{} the properties of the X-ray halo are those of the halo of a $\sim$L$^*$ galaxy, with stellar mass M$_{*} = 10^{10.53}$ M$_{\odot}$ virial radius $R_{vir} \simeq 195$ kpc, virial temperature T$_{vir}\simeq 10^6$ K, and halo mass
  M$_h = 10^{12.1}$ M$_{\odot}$;
 \item{} we estimate the mass of the cool-CGM phase of the X-ray halo to be about half the average stellar mass of the three galaxies that host it, i.e. M$_{cool-CGM} = 10^{10.27}$ M$_{\odot}$,
  which leaves a missing baryon mass in the system M$_{missing} \simeq 1.45\times 10^{11}$ M$_{\odot}$;
\item{} our line of sight intercepts the X-ray halo at weighted-average projected distance of $\rho = 115$ kpc, i.e. $\simeq 0.6$ R$_{vir}$ and, in a spherical configuration, has a pathlength of $L = 315$ kpc
  through the halo, along which we estimate equivalent-hydrogen column densities of the \ovii-bearing gas, at the average $<b_{OVI}> = 68\pm 10$ km s$^{-1}$, which are virtually independent on whether
estimated from the best-fitting \ovii (and upper limit on \oviii) EWs derived from the stacked X-ray-LLS or FUV-LLS spectra, i.e. logN$_H^{X-ray-LLS}$(in cm$^{-2}$)$=(19.86–20.51) (Z/0.3Z_{\odot})^{-1}$ 
  or logN$_H^{FUV-LLS}$(in cm$^{-2}$)$=(20.02–20.59) (Z/0.3Z_{\odot})^{-1}$; 
\item{} by assuming a spherical geometry and a $\beta$ density-profile for the X-ray halo, we derive hot-CGM masses again largely independent on the stacking methodology and in the ranges  M$_{hot-CGM}^{X-ray-LLS} = (1-1.7)
  \times 10^{11} (Z/0.3Z_{\odot})^{-1}$ M$_{\odot}$ or M$_{hot-CGM}^{FUV-LLS} = (1.3-1.6) \times 10^{11} (Z/0.3Z_{\odot})^{-1}$ M$_{\odot}$, corresponding to missing-baryon fractions $\xi_b^{X-ray-LLS} = \frac{M_{hot-CGM}}{M_{missing}} =
  (0.7-1.2) (Z/0.3Z_{\odot})^{-1}$, $\xi_b^{FUV-LLS} = (0.92-1.02) (Z/0.3Z_{\odot})^{-1}$, and temperatures in the intervals logT(in K)$^{X-ray-LLS} \simeq 5.78-6.28$, logT(in K)$^{FUV-LLS} \simeq 5.82-6.2$, both comprising the X-ray-halo virial temperature 
$T_{vir} \simeq 10^6$ K.
\end{itemize}

Our findings imply that virtually all the baryons that were still missing in typical L$^*$ galaxies, can now be accounted for by the hot-CGM gas. This has important consequences for our understanding of the
galaxy-CGM and galaxy-IGM feedback processes throughout the universe lifetime, suggesting that accretion in these galaxies mostly occurred in hot-mode and at the rate given by the cosmological
baryon fraction. Feedback from internal activity was efficient in metal-polluting the CGM and perhaps hampering its cooling, but not sufficient to expel a significative fraction of the baryons
beyond R$_{vir}$.

{\color{black} 
\section{Scripts and Code Availability}
The paper makes use of publicly available codes for spectral fitting (i.e. {\em Sherpa}) and custom-made Fortran90 and SuperMongo routines for: (a) the curve-of-growth analysis (Fortran90; first developed, used and published in \cite{Nicastro99} and then also used
in e.g. \cite{Nicastro05,Williams05,Nicastro18}); (b) the X-ray-halo mass calculation via $\beta$-profile functions (Fortran90); (c) the de-redshifting and regridding over a common spectral grid of individual source and background spectra and best-fitting continuum
models (Fortran90); and (d) to generate figures (SuperMongo). All custom-made routine and software shall be made available by the corresponding author, upon request.
}

\begin{acknowledgments}
  We thank the anonymous referees for the useful comments and suggestions, which significantly helped improving the paper. 
F.N., S.B., M.B, A.B., A.D.R., C.F., E.P. and L.Z. acknowledge support from INAF-PRIN grant “A Systematic Study of the largest reservoir of baryons and metals in the universe: the circum-galactic medium of galaxies”
(\#1.05.01.85.10). F.N., E.P. and L.Z. also acknowledge support from the EU grant AHEAD-2020 (GA \#871158). T.F. acknowledges support by the National Key R\&D Program of China No. 2017YFA0402600, NSFC grants No. 11890692
and 12133008 and the science research grants from the China Manned Space Project with NO. CMS-CSST-2021-A04. 
\end{acknowledgments}

\vspace{5mm}
\facilities{\xmm(RGS), \chandra(LETG), HST(COS)}

\appendix
\section{Selection of the X-ray-Halo Sample and Data Processing} \label{appsec:sample-selection}
We cross-correlated the \xmm RGS and \chandra High Resolution Camera (HRC, \cite{Murray00}) LETG archives with the LLS samples of \cite{Lehner13} and \cite{Prochaska19}, consisting of 30 quasars observed 
with the HST-COS crossing LLSs with moderate HI column density ($16.2 \le$logN$_{HI}$(in cm$^{-2}$)$\le 19$) and associated low- and moderate-ionization metal absorbers \citep{Fox13,Lehner13}. 
We found 11 matches. Seven of these have more than one \xmm RGS public observations, while the remaining four objects have been observed only once and with very short exposures (16-28 ks each) and were removed from the sample. 
{\color{black} Of the 7 targets with multiple RGS spectra (namely PG~1407+265, PG~1116+215, PG~1216+069, PKS~0312-77, PG~1634+706, PKS~0405-123 and PHL~1811), PG 1216+069 (total exposure of 100 ks) is a calibration source and 
was observed several arcminutes off the aimpoint, thus with a severely degraded spectral resolution. This source was also removed from the sample. 
We downloaded all the available RGS data of the remaining 6 targets and reprocessed them with the latest version of the \xmm Science Analysis Software (SAS v. 20.0.0) and calibration, to produce a final co-added 
RGS1+RGS2 spectrum of each target. This was done by first using the SAS tool {\em rgsproc} withh all default parameters, except {\em keepcool}, set to {\em no}, and {\em witheffectiveareacorrection}, set to {\em yes}. This produced all standard
products, including RGS1 and RGS2 source and background spectra, response matrices and background lightcurves. For each observations, background lightcurves were checked for background flares caused by high fluxes of soft protons
hitting the detectors, normally during the first or last parts of the observations. About half of the RGS observations were contaminated by background flares (a rate of $\ge 0.1$ cts s$^{-1}$ in the standard background extraction regions) for 3-5\% of
their total exposures, and these time-intervals were filtered out from the affected observations by re-running {\em rgsproc} with non-standard GTI-filters.
Finally, spectra of the same target from single RGS1 and RGS2 observations were coadded by using the SAS tool {\em rgscombine}, which also produces averaged responses for the final coadded spectrum. 
We checked these coadded source and background spectra of our targets in spectral regions close to the LLS-frame \ovii K$\alpha$ transition, and verified that sources counts largely dominate all spectra in these regions, displaying 
extraction-region-normalized source/background ratios per spectral resolution elements $\ls 3$}.

\noindent
Two of the six objects of our \xmm sample have also multiple \chandra HRC-LETG observations publicly available. We downloaded these \chandra observations and reprocessed them with the latest version of the 
\chandra Interactive Software of Observations (CIAO v.4.14), \cite{Fruscione06}, to produce final total LETG spectra of these targets. {\color{black} We did this by first using the standard {\em Ciao} script {\em chandra\_repro} on all
HRC-LETG observations of our sample, to produce negative- and positive-oder source and background spectra (and convolved $1-6$-order {\em effective-area $\times$ photon-redistribution} responses: HRC-LETG does not resolve orders)
of each observations.
Source and background spectra of the same target, were then coadded by exploiting the {\em Ciao} tool {\em combine\_grating\_spectra}}. 

Finally, we imposed the two following selection criteria to the targets of our sample: 
{\color{black} (a) that the LLSs detected along their lines of sight have been confidently associated to Milky-Way-like galaxies, down to sub-L$^*$ luminosities (e.g. \cite{Lehner13}), and (b) that their total RGS and/or LETG spectra
have SNRE$>4$ in the continuum adjacent the relevant lines, which, in turn, guarantees the use of Poissonian statistics on the data and the $\ge 90$\% sensitivity to \ovii K$\alpha$ absoprtion lines with LLS-frame EW$\gs 10-50$ m\AA\
in the redshift range $z\simeq 0.1-0.6$ (typical of intervening hot-CGM Milky-Way-like halos in hydrodynamical simulations, e.g. \cite{Wijers20}) 
The adoption of both criteria guarantees that the signal in the stacked (or simultaneously fit) spectra is not washed out by the absence (either because of no clear galaxy-LLS associaton or because of the possible intrinc absence of hot 
gas associated to the LLS along the particular line of sight) of high-ionization metal X-ray absorption in any of the stacked spectra}. 
Only 4 targets passed our second selection criterion (namely, PG~1407+265, PKS~0405-123, PG~1116+215 and PG~1216+069) and only three of these four lines of sight (PG~1407+265, PKS~0405-123 and PG~1116+215) interecept LLSs 
that have been confidently associated to $\sim L^*$ galaxies. For the LLS along the line of sight to PG~1216+069, the closest galaxy, down to 0.1 L$^*$ luminosities, lies at an inpact parameter of 3.2 Mpc from the LLS \citep{Lehner13} and,
consistently, the RGS spectrum of this target shows no hint of LLS-associated \ovii$_{K\alpha}$ absorption, suggesting that the LLS along this line of sight is probably imprinted by an intervening Lyman-$\alpha$-forest filament rather than
the cool-CGM of a intervening galaxy (this target was indeed also removed from the LLS sample of \cite{Lehner13} in their subsequent works, e.g. \cite{Berg23}). 

Thus, our final X-ray-halo sample consists of 3 targets, with a total of 5 X-ray spectra: 3 \xmm-RGS and 2 \chandra-LETG.
 
Table \ref{apptab:xspec} summarizes the properties of the targets of our X-ray-halo sample and their X-ray spectra. The last row of Table \ref{apptab:xspec} lists the total available X-ray exposure and SNREs (added 
in quadrature). 
\begin{table}[ht!]
\centering
\caption{ \it Properties of the X-ray Spectra of the X-ray-Halo Sample}
\vskip 0.1 in
\begin{tabular}{|l|l|c|c|c|c|c|}
\hline
QSO & LLS & $^a$$z_{em}$ & Exposure (ks) & Exposure (ks) & $^b$SNRE & $^b$SNRE \\
    & & & RGS1+RGS2 & LETG & RGS1+RGS2 & LETG \\
\hline
\hline
PG~1407+265  & 1 & 0.94 & 213 & NA & 4.7 & NA \\
PKS~0405-123 & 2 & 0.5726 & 1402 & 376 & 17.2 & 6.1 \\
PG~1116+215 & 3 & 0.1756 & 776 & 355 & 20.9 & 8.1 \\
\hline
\multicolumn{7}{|c|}{Stacked X-ray-halo Spectrum} \\
\hline
Tot \& Averages & NA & NA & 2391 & 731 & 27.5 & 10.1 \\
\hline
\end{tabular}

$^a$ redshift of the background quasars. 
$^b$ Signal-to-Noise per Resolution Element {\color{black} at $\lambda_{RF} = 21.6$\AA\ (in the LLS-frame)}, where the \ovii K$_{\alpha}$ transition lies. \\
\label{apptab:xspec}
\end{table}

\section{Properties of the LLS and the X-ray Halo} \label{appsec:xray-halo_properties}
Table \ref{apptab:xray-halo} lists the properties of the LLS–galaxy associations relevant to this work: namely the stellar-mass M$_*$, halo mass M$_h$ (defined as M$_{200}$
\footnote{the mass embedded in a sphere with radius R$_{200}$}
) and virial radius R$_{vir}=$R$_{200}$ of the galaxies, together with the impact parameter $\rho$, the metallicity of the cool-CGM in the LLSs and the column and Doppler parameter of the LLS-associated \ovi absorbers. 

Virial radii of the galaxy-associations are reported in the literature for all three LLSs of our X-ray halo \citep{Berg23,Keeney17,Burchett19}, while the halo mass is reported only for the LSS\#2 along 
the sightline to PKS 0405-123 \citep{Berg23}. For LLS\#2 Berg and collaborators (2023) estimate M$_h$ via the stellar-mass--halo-mass relation, as in \cite{Rodriguez-Puebla17}, and then derive R$_{vir}$
via the relationship M$_h = 4/3 \pi R_{vir}^3 \Delta_{200} \rho_c (z)$, where $\rho_c (z) = (3H_0^2/8\pi G)[\Omega_m(1+z)^3 + \Omega_{\Lambda}]$ is the critical density of the universe at redshift z. 
For the other two LSSs R$_{vir}$ is derived in \cite{Keeney17} and \cite{Burchett19} through abundance-matching, i.e. via the galaxy’s optical luminosity, by matching an observed galaxy luminosity function with a theoretical
halo-mass function. For these two galaxy-associations, we derive M$_h$ from R$_{vir}$, again as in \cite{Berg23}, through the relationship M$_h = 4/3 \pi R_{vir}^3 \Delta_{200} \rho_c (z)$.
Finally, the last row of Table \ref{apptab:xray-halo} lists the property of the X-ray halo that we use in this work. These are derived (all but the Doppler parameter of \ovi, which is further weighted by the column density of \ovi)
by weighting the quantities in rows 1–3 by the statistical significance of the \ovii K$\alpha$ lines in the spectra of our three targets (i.e. $\sigma^i_{OVII} = 1.7$, 2.8 and 2.8 for PG~1407+265,
PKS~0405-123 and PG~1116+215, respectively: Tab. \ref{tab:single-spectra_linepars}), and averaging them. 

\begin{table}[ht!]
\centering
\caption{ \it Properties of the LSS and the X-ray Halo}
\vskip 0.1 in
\begin{tabular}{|l|c|c|c|c|c|c|c|c|}
\hline
QSO (LLS \#) & $z_{LLS}$ & M$_*$ & M$_h$ & R$_{vir}$ & $\rho$ & [X/H] & logN$_{OVI}$ & b$_{OVI}$ \\
  & & (in logM$_{\odot}$) & (in logM$_{\odot}$) & (in kpc) & (in kpc) & & (in cm$^{-2}$) & (in km s$^{-1}$) \\
\hline
\hline
PG~1407+265 (\#1) & 0.6828 & $^a$10.9  & 12.4 & $^a$220 & $^a$91 & $^b$-1.66 & $^c$$13.99 \pm 0.06$ & $^c$$28 \pm 10$ \\
PKS~0405-123 (\#2) & 0.1672 & $^d$10.3  & $^d$11.9 & $^d$183 & $^d$117 & $^b$-0.29 & $^c$$14.59 \pm 0.05$ & $^c$$78 \pm 10$ \\
PG~1116+215 (\#3) & 0.1385 & $^e$10.3 & 11.9  & $^f$192 & $^g$127 & $^b$-0.56 & $^c$$13.85 \pm 0.05$ & $^c$$47 \pm 10$ \\
\hline
\multicolumn{9}{|c|}{X-ray Halo} \\
\hline
Weighted Averages & 0.276 & 10.53 & 12.1 & 195 & 115 & -0.512 & $14.30 \pm 0.05$ & $68\pm 10$ \\
  \hline
\end{tabular}

$^a$\cite{Burchett19}. $^b$\cite{Wotta19}. $^c$\cite{Fox13}. $^d$\cite{Berg23}. $^e$Assumed to be the same as PKS~0405-123, given the same halo mass. $^f$\cite{Keeney17}. $^g$\cite{Lehner13}. 
\label{apptab:xray-halo}
\end{table}

\section{Uncertainties in the HRC-LETG and RGS wavelength scales.} \label{appsec:disp-rels}
{\color{black} Fig. \ref{fig:individual-contours} and Table \ref{tab:single-spectra_linepars} show that the best-fit LLS-frame centroids of the putative \ovii K$\alpha$ lines of virtually all the available RGS and LETG spectra of the X-ray halo, 
are offset from the expected $\lambda = 21.6$ \AA\ position, though for four out of the five spectra, by less than the LSF FWHM of the spectrometer (70 and 50 m\AA\ for the RGS and the LETG respectively).
The exception is the line detected at a $2\sigma$ confidence level in the HRC-LETG spectrum of PG~1116+215 (made up of 11 different exposures), which is offset by 120 m\AA\ (i.e. 2.4$\times$FWHM$_{LETG-LSF}$) from the
line rest-frame wavelength. The same line is seen at a similar statistical confidence level in the RGS spectrum of the same target (1.8$\sigma$) and its position is shifted in the same drection but only by 60 m\AA, i.e. 0.86$\times$FWHM$_{RGS-LSF}$
from both the line rest-frame position and the centroid of the line in the LETG spectrum.} 

The large displacement of the line position in the \chandra spectrum of PG~1116+215, could at least partly be due to the large systematic uncertainties that affect the HRC-LETG dispersion relationship. 
According to the \chandra HRC-LETG calibrations, the wavelength scale of the HRC-LETG spectrometer suffers uncertainties of up to 50 m\AA\ (about 1000 km s$^{-1}$ at the wavelength of the FeXVII line where such
uncertainty has been evaluated), due to the non-linearity of the dispersion relationship, which, in turn, is due to the non-linear imaging distortions of the HRC-S detector
\footnote{\url{https://cxc.cfa.harvard.edu/cal/letg/Corrlam/}}
. Such distortions are randomly spaced in wavelengths across the entire spectrum and therefore cannot be calibrated based on the presence of strong lines with known positions in different regions of the same 
spectrum (which, in any case, are not present in the LETG spectrum of PG~1116+215). 
Moreover, the 1000 km s$^{-1}$ calibration uncertainty quoted above has been derived for the strong FeXVII ($\lambda \simeq 15$ \AA) emission line of a very bright X-ray star (Capella). 
The velocity-space uncertainty in the aspect reconstruction of the dispersion-relation may be larger for fainter lines, especially in absorption against relatively low-flux continua, and could propagate randomly 
when adding together different low-exposure observations of the same source, {\color{black} with the net effect of shifting the unresolved-line centroid even beyond one nominal resolution element. 

To investigate this further, we decided to use two samples of absorption lines detected in the X-ray spectra of extragalactic targets, namely the Galactic \ovii K$\alpha$ and K$\beta$ ($\lambda=21.6$ and 18.63 \AA, respectively) 
and \oi and \oii K$\alpha$ ($\lambda=23.52$ and 23.35 \AA, respectively) lines seen unbiquitously in spectra of both Galactic and extragalactic targets with sufficient SNRE (\cite{Nicastro16a, Nicastro16b}). 
The sample of \ovii lines is that of \cite{Nicastro16a}, extracted from RGS spectra with SNRE$>10$ at 21.6 \AA, and contains 34 \ovii K$\alpha$ lines, for 16 of which also the associated K$\beta$ line is detected. 
The sample of \oi and \oii K$\alpha$ lines, instead, is that of \cite{Nicastro16b}, extracted from HRC-LETG spectra with SNRE$>3$ at 23.5 \AA, and contains 11 \oi K$\alpha$ lines, for 10 of which also the associated \oii K$\alpha$ 
line is detected. 
Fig. \ref{appfig:Galactic_OVII-OI_Centroid-Distribution} shows the probability-density distributions (in bins of 500 km s$^{-1}$) of the offsets between the best-fitting \ovii K$\alpha$, K$\beta$ (RGS lines: black solid histogram) and 
\oi, \oii K$\alpha$ (HRC-LETG lines: cyan solid histogram) line-centroids and the rest-frame wavelengths of these transitions, together with their cumulative distributions (dashed black and cyan histograms, respectively).
The inset of Fig. \ref{appfig:Galactic_OVII-OI_Centroid-Distribution} shows the single \ovii K$\alpha, \beta$ (black points and 1$\sigma$ statistical-only errorbars) and \oi, \oii K$\alpha$ (cyan points and 1$\sigma$ statistical-only errorbars)
line-centroid offsets, as a function of the SNRE in the spectra. 
Both in the main figure and in the inset, colored points and 1$\sigma$ statistical-only error-bars are the measured \ovii K$\alpha$ LLS-frame line-centroid offsets in the five spectra of our targets (orange: PG~1407+265; green:
PKS~0405-123 and violet: PG~1116+215). 

Both distributions are broad, spanning a range of about 2000 km $^{-1}$ in line-centroid offsets, and, in the adopted binning scheme, have $FWHM_{RGS} \simeq 1000$ km s$^{-1}$  and $FWHM_{LETG} \simeq 1500$ km s$^{-1}$,
about 1$\times$ and 2.2$\times$ the nominal RGS and LETG LSF FWHMs at $\lambda = 21.6$ \AA, respectively.
In these paper we assume, as 1$\sigma$ (statistical plus systematic) errors for the best-fitting line centroids in the two instruments, the Gaussian-equivalent standard deviation of the distributions, i.e. $\Delta\lambda_{1\sigma} = \pm$
FWHM$/(2\sqrt{2ln2})$. 
The  centroid of the \ovii K$\alpha$ line in the HRC-LETG spectrum of PG~1116+215 is marginally consistent, within its 1$\sigma$ statistical-only error, with both the \ovii K$\alpha$ line seen in the RGS spectrum of the 
same target (compare the two violet errorbars in the main panel of Fg. \ref{appfig:Galactic_OVII-OI_Centroid-Distribution}) and the negative tail of the observed HRC-LETG \oi $K\alpha$ centroid distribution (sampled with a
probability of about 5\% in our distribution, i.e. one out of the 21 \oi, \oii K$\alpha$ HRC-LETG line centroid measurements: Fig. \ref{appfig:Galactic_OVII-OI_Centroid-Distribution}). 
\begin{figure}[ht!]
\plotone{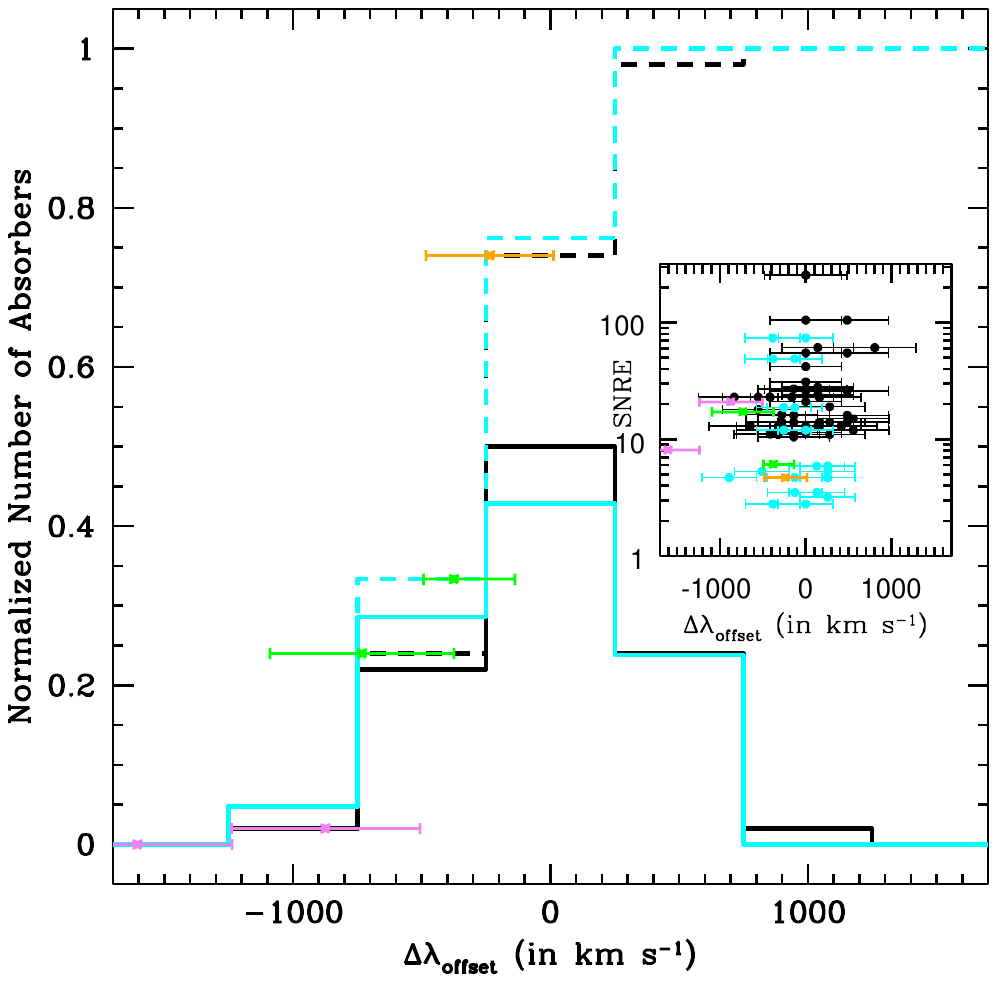}
\caption{Probability-density distributions of the offsets between the best-fitting line centroids of the \ovii K$\alpha$, K$\beta$ (RGS lines: black solid histogram) and \oi, \oii K$\alpha$ (HRC-LETG lines: cyan solid histogram) absorbers from the
  samples of \cite{Nicastro16a} and \cite{Nicastro16b}, respectively, and the rest-frame wavelengths of these transitions, together with their cumulative distributions (dashed black and cyan histograms, respectively). The inset shows the single
  \ovii K$\alpha, \beta$ (black points and 1$\sigma$ statistical-only errorbars) and \oi, \oii K$\alpha$ (cyan points and 1$\sigma$ statistical-only errorbars) line-centroid offsets, as a function of the SNRE in the spectra.}
\label{appfig:Galactic_OVII-OI_Centroid-Distribution}
\end{figure}

The breadths of the RGS and LETG line-centroid offset distributions are not due to rigid shifts of the dispersion relationship of the two spectrometers, from observation to observation. 
Indeed, these distributions become even broader when the offset between the observed and expected relative positions of two known lines is considered. Fig. \ref{appfig:Galactic_kakb-OI-OII_Delta-Distribution} shows
the probability-density (solid histograms) and cumulative (dashed histograms) distributions of the offsets (in km s$^{-1}$) between the \ovii K$\alpha$, K$\beta$ and the \oi, \oii K$\alpha$ line-centroid differences measured in our
RGS (black histograms) and HRC-LETG (cyan histograms) samples, respectively, and the corresponding expected line-centroid differences.
As in Fig. \ref{appfig:Galactic_OVII-OI_Centroid-Distribution}, the inset shows the data-points from our samples (black from the RGS and cyan from the HRC-LETG) as a function of the SNRE at the relevant wavelengths.
Both in the main figure ad in the inset, the orange point with its 1$\sigma$ statistical error-bars, is the offset between the relative positions of the \ovii K$\alpha$ and K$\beta$ lines measured in the stacked spectrum
of the X-ray halo obtained by rigidly shifting each X-ray spectrum to the exact FUV-LLS redshifts (i.e. right panels of Fig. \ref{fig:stacked-spectrum} and \ref{fig:contours-stacked}), and their expected relative position.
The observed offset is fully consistent with both the RGS and HRC-LETG distributions.

\begin{figure}[ht!]
\plotone{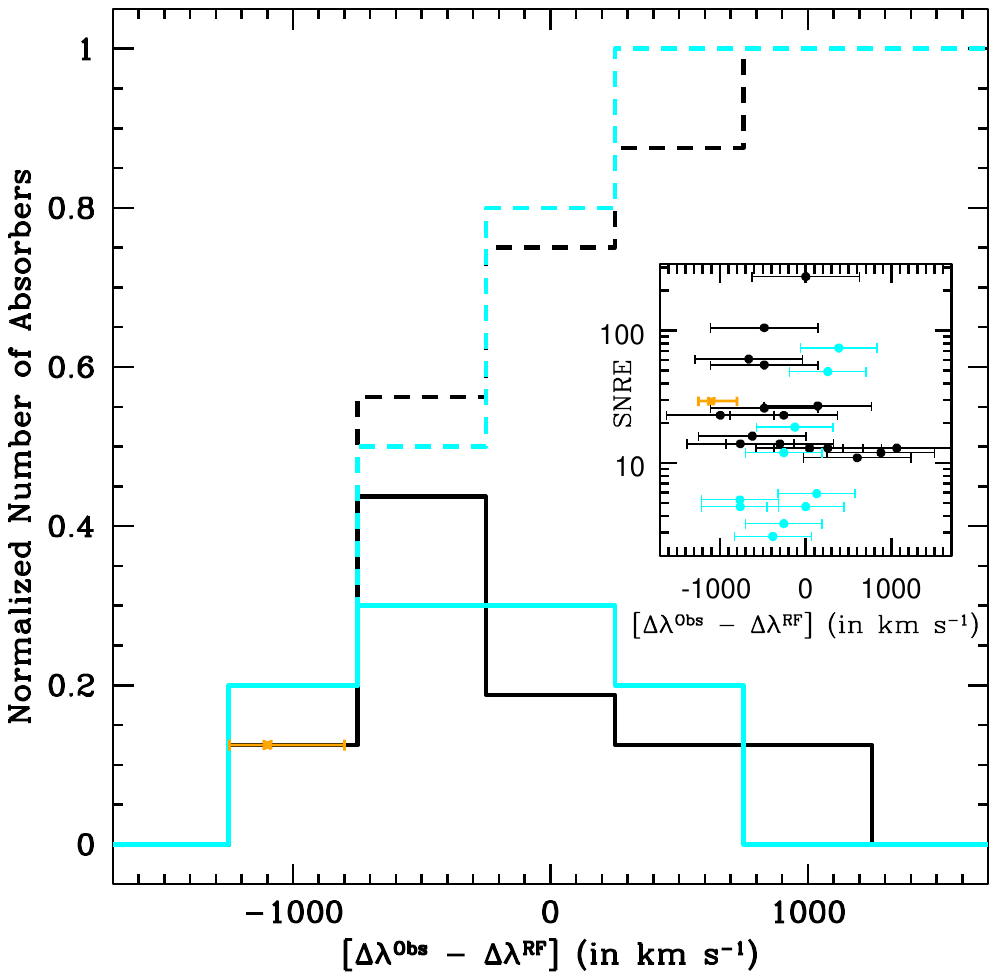}
\caption{As Fig. \ref{appfig:Galactic_OVII-OI_Centroid-Distribution} but for the offsets (in km s$^{-1}$) between the \ovii K$\alpha$, K$\beta$ and the \oi, \oii K$\alpha$ line-centroid differences measured in our
RGS (black histograms) and HRC-LETG (cyan histograms) samples, respectively, and the corresponding expected line-centroid differences.}
\label{appfig:Galactic_kakb-OI-OII_Delta-Distribution}
\end{figure}
}

\section{Ion Fractions in CIE and PIE gas} \label{appsec:ion-fractions}
Figure \ref{appfig:ion-fractions} shows the fractional abundances of the ions \ovi (green), \ovii (blue) and \oviii (brown) as a function of temperature (left panel) and hydrogen density (right panel), in gas in collisional 
ionization equilibrium (CIE; left panel) and photoionized by the average meta-galactic radiation field at the redshift of the X-ray halo (PIE, \cite{Nicastro16b}; right panel), respectively. 

In CIE gas (left panel) \ovii is effectively the only populated ion of oxygen at logT(in K)$\simeq 5.7-6.1$ (the virial temperature range for halo masses in the range of M$_h 
\simeq 10^{11.7-12.3}$ M$_{\odot}$, at the X-ray halo redshift $z=0.276$), while \ovi and \oviii fractions peak, respectively, at the opposite extremes of the considered temperature range, namely logT(in K)$\simeq 5.5$ and
T(in K)$\simeq 6.4$, and reach maximum abundances of only 20 and 40\%.
Thus virialized gas in typical L$^*$ galaxy's halos can efficiently produce \ovii but only small fractions of \ovi and \oviii, the first still detectable in current FUV spectra of bright background targets. 

On the contrary, PIE gas illuminated by the meta-galactic radiation can produce sizeable fractions of \ovii only at typical IGM densities $n_H \ls 10^{-5.3}$ cm$^{-3}$), while \ovi can still be moderately populated and detectable at
typical galaxy-halo densities $n_H\simeq 10^{-4}-10^{-5}$ cm$^{-3}$. 

\begin{figure}[H]
\plotone{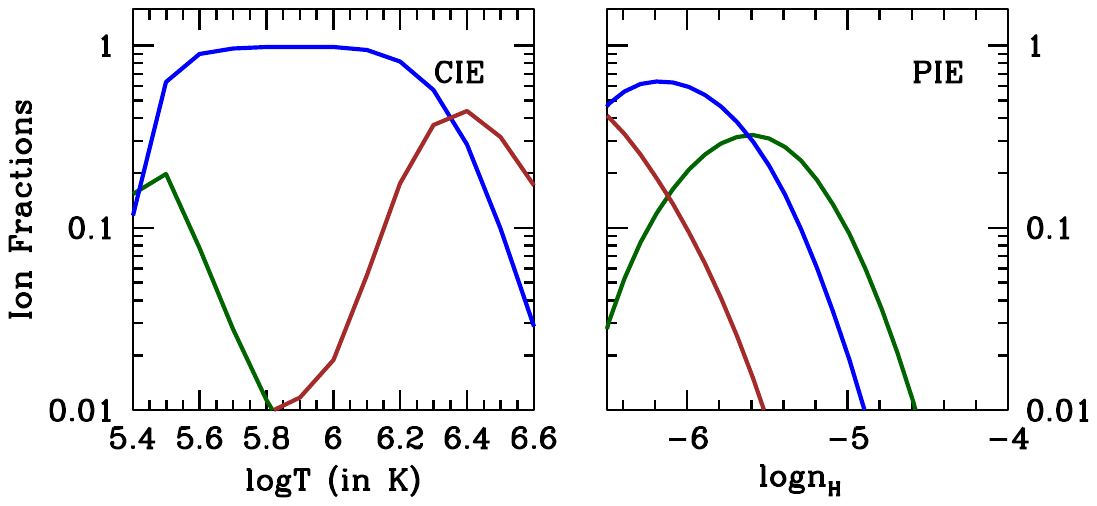}
\caption{Fractional abundances of \ovi (green), \ovii (blue) and \oviii (brown) as a function of temperature (left panel) and hydrogen density (right panel), in CIE (left) and PIE gas illuminated 
by the average meta-galactic radiation field at $z=0.276$ (right).} 
\label{appfig:ion-fractions}
\end{figure}

\bibliography{biblio}{}
\bibliographystyle{aasjournal}

\end{document}